\documentclass[sigplan, nonacm]{acmart}

\usepackage{subcaption}
\usepackage{algorithm}
\usepackage{balance}
\usepackage{algpseudocode}
\usepackage{placeins}
\usepackage{xcolor}
\usepackage{listings}
\lstdefinestyle{mystyle}{
    backgroundcolor=\color{lightgray!20},   
    commentstyle=\color{orange},
    keywordstyle=\color{blue},
    numberstyle=\tiny\color{gray},
    stringstyle=\color{orange},
    basicstyle=\ttfamily\fontsize{7.8}{9}\selectfont,
    breaklines=true,
    breakindent=0pt,
    breakautoindent=false,
    captionpos=b,
    numbers=left,
    numbersep=5pt,
    showspaces=false,
    showstringspaces=false,
    showtabs=false,
    columns=flexible,
    tabsize=1,
    language=Python
}

\newcommand{\showcomments}{no}
\newcommand{\NVLdomain}{\emph{scale-up domain} }

\newcommand\daiyaan[1]{
    \ifthenelse{\equal{\showcomments}{yes}}{\textcolor{red}{[daiyaan: #1~]}}{\ignorespaces}
}
\newcommand\greg[1]{
    \ifthenelse{\equal{\showcomments}{yes}}{\textcolor{green}{[greg: #1~]}}{\ignorespaces}
}

\newcommand\dheevatsa[1]{
    \ifthenelse{\equal{\showcomments}{yes}}{\textcolor{blue}{[dheevatsa: #1~]}}{\ignorespaces}
}
\newcommand\ahmet[1]{
    \ifthenelse{\equal{\showcomments}{yes}}{\textcolor{violet}{[ahmet: #1~]}}{\ignorespaces}
}
\newcommand\ankit[1]{
    \ifthenelse{\equal{\showcomments}{yes}}{\textcolor{cyan}{[ankit: #1~]}}{\ignorespaces}
}

\acmConference[]{}{}{}
  
\begin{document}
\title{Nonuniform-Tensor-Parallelism: Mitigating GPU failure impact for Scaled-up LLM Training}

\author{Daiyaan Arfeen}
\authornote{Work done during internship at NVIDIA.}
\affiliation{%
  \institution{Carnegie Mellon University}
  \city{Pittsburgh, PA}
  \country{USA}
}

\author{Dheevatsa Mudigere}
\affiliation{%
  \institution{NVIDIA}
  \city{Santa Clara, CA}
  \country{USA}
}

\author{Ankit More}
\affiliation{%
  \institution{NVIDIA}
  \city{Santa Clara, CA}
  \country{USA}
}

\author{Bhargava Gopireddy}
\affiliation{%
  \institution{NVIDIA}
  \city{Santa Clara, CA}
  \country{USA}
}

\author{Ahmet Inci}
\affiliation{%
  \institution{NVIDIA}
  \city{Santa Clara, CA}
  \country{USA}
}

\author{Gregory R. Ganger}
\affiliation{%
  \institution{Carnegie Mellon University}
  \city{Pittsburgh, PA}
  \country{USA}
}

\renewcommand{\shortauthors}{Daiyaan Arfeen et al.}

\begin{abstract}
LLM training is scaled up to 10Ks of GPUs by a mix of data- (DP) and model-parallel (MP) execution.
Critical to achieving efficiency is tensor-parallel (TP; a form of MP) execution within tightly-coupled subsets of GPUs, referred to as a \emph{scale-up domain}, 
and the larger the scale-up domain the better the performance.
%and the larger the TP degree (number of GPUs used together via TP) the higher the performance.
New datacenter architectures are emerging with more GPUs able to be tightly-coupled in a scale-up domain, such as moving from 8 GPUs to 72 GPUs connected via NVLink.
%, with tensor-parallelism (TP) being a cModel-parallelism is a key technique for scaling up LLM training specifically Tensor-Parallelism (TP), and to enable increasing TP new datacenter architectures are emerging with increasing number of compute nodes that are tightly coupled together in a \NVLdomain and several such blocks are scaled out to form the larger cluster. 
%Google's TPU pods, GPUs connected using high-bandwidth scale-up interconnect (NVIDIA's NVLink, AMD's infinity fabric link), which have been growing in size with subsequent generations are examples of this system design trend.
Unfortunately, larger scale-up domains increase the blast-radius of failures, with a failure of single GPU potentially impacting TP execution on the full scale-up domain, which can degrade overall LLM training throughput dramatically.
%can degrade dramatically for large-model training jobs. 
With as few as 0.1\% of GPUs being in a failed state, a high TP-degree job can experience nearly 10\% reduction in LLM training throughput.
%(reducing throughput by 10\%); at pretraining scales this amounts to thousands of GPUs lost.
%Furthermore, model-parallelism mapped to such scale-up blocks are constrained to exact dimension. For instance, with fixed TP degrees requiring strict NVLink domain sizes. 
%To overcome both these challenges, 

We propose nonuniform-tensor-parallelism (NTP) to mitigate this amplified impact of GPU failures.
In NTP, a DP replica that experiences GPU failures operates at a reduced TP degree, contributing throughput equal to the percentage of still-functional GPUs.
We also propose a rack-design with improved electrical and thermal capabilities in order to sustain power-boosting of scale-up domains that have experienced failures; combined with NTP, this can allow the DP replica with the reduced TP degree (i.e., with failed GPUs) to keep up with the others, thereby achieving near-zero throughput loss for large-scale LLM training.
%Through a hierarchical sharding technique and overlapping re-sharding with backward computation and parameter-synchronization, NTP introduces near-zero overhead and largely eliminates failure amplification. 
\end{abstract}

\maketitle

\daiyaan{TODO:
\begin{itemize}
    \item discussion failure studies
    \item dlsim breakdown for TPs
    \item implementation details (if any missing e.g. code snippets) 
    \item broader discussion on heterogeneous parallelism 
    \item cite larger NVL domains presented at GTC 
\end{itemize}
}

\section{Introduction}

Training large-language models (LLMs) and other generative AI models involves thousands of GPUs\footnote{We use ``GPU'' to refer to AI accelerators generally, such as GPUs, TPUs~\cite{tpuv4}, and Trainium~\cite{trainium}} used in parallel for weeks.
Generally speaking, the models are huge and a mix of parallelization approaches is required to achieve scale and efficiency.
Data parallel (DP) execution is used to orchestrate multiple sets of GPUs (128--2K) that each process input samples for one replica of the model, which is generally too large to fit in the high-bandwidth memory (HBM) of a single GPU.
The set of GPUs for each replica uses pipeline parallelism (PP) to divide model layers across subsets of GPUs (e.g., 8--16) that use so-called tensor parallelism (TP; tightly connected division of each layer) for each pipeline stage worth of layers.

GPU efficiency for large-scale LLM training can be higher when the TP degree (number of GPUs used together via TP) is larger.
But TP only works well with GPUs that are tightly connected via a high-bandwidth interconnect like NVIDIA's NVLink~\cite{nvlink} or AMD's infinity fabric link~\cite{infinity-fabric} or Google's TPU Pods~\cite{tpu-pod}, limiting the TP degree in practice.
For example, current NVLink configurations connect up to 8 GPUs.
Fortunately, the \emph{scale-up domain} (the number of tightly connected GPUs) has been growing, with up to 72 being released into production~\cite{gb200-nvl72}, promising significant efficiency improvements for large-scale LLM training using 10s of 1000s of GPUs.

Larger scale-up domains, however, lead to larger numbers of GPUs being affected by a single GPU failure---the entire TP execution spread across all of the GPUs in a single scale-up domain becomes unavailable, in turn affecting the entire data-parallel replica.
This amplification of failure impact can greatly reduce overall training efficiency.
For example, for TP degree 64 (TP64), just 0.1\% of the GPUs being in a failed state can lead to nearly 10\% of the GPUs allocated for training not contributing to training throughput.
For a 32K-GPU training job, this would be $\approx$3000 GPUs.

This paper proposes \emph{Nonuniform Tensor Parallelism (NTP)} as a resilient distributed training approach to address this problem.
In NTP, a DP replica that experiences one or more GPU failures will reconfigure to operate at a lower TP degree (e.g., TP62 if two of the 64 GPUs in the given scale-up domain have failed) than the other fully healthy DP replicas.
Naturally, since the reconfigured DP replica has fewer GPUs for the same model, its throughput would be expected to drop.
To avoid a persistent straggler that slows bulk-synchronous execution, the simplest solution is to reduce the number of input samples for the reconfigured replica, which minimizes the impact on training throughput when using traditional machines.
Near-zero throughput loss becomes feasible, however, with an alternate rack-design that allows power-boosting in the scale-up domain with the failed GPU---speeding up just the one reconfigured TP instance can allow it to keep up with the others at near-zero average energy usage increase.

Realizing NTP requires addressing the more complex communication patterns involved in synchronizing parameter updates among DP replicas.
Specifically, since the replicas are no longer identical, all-reduce cannot involve just the corresponding machine from each of the DP replicas---parameters are distributed among machines differently in the reconfigured scale-up domains with failed GPU(s).
We address this by resharding gradients among GPUs in DP replicas when some peer GPUs are in reconfigured TP instances, before and after using traditional all-reduce among those GPUs.
We avoid significant performance degradation by aggressively overlapping resharding with backward computation and parameter synchronization, with measurements of our prototype implementation showing $<$1\% slowdown (from resharding) for healthy replicas. Evaluations with a well-calibrated simulator show the resilience of our system: ~3\% throughput loss at very high cluster failure fractions for our software solution and <1\% throughput loss when power-boosting partially failed scale-up domains.

This paper makes four primary contributions:
(1) it identifies and quantifies the high failure-amplification effects of increasing scale-up domains in LLM training;
(2) it introduces nonuniform-tensor-parallelism (NTP) - a resilient distributed training approach to mitigate such failure-amplification;
(3) it shows how NTP can be implemented efficiently and with minimal communication overhead;
(4) it shows how new rack designs that allow small subsets of GPUs to be power-boosted can enable near-zero LLM training throughput loss in the face of GPU failures.

\section{Background and Motivation}
Here we will describe distributed LLM training workloads and how they are parallelized; we will show why tensor-parallelism is often preferred despite inducing high-volume blocking communication patterns. We will then show how increasing \NVLdomain size, which increases the degree of tensor-parallelism that can be used, increases the failure amplification/blast-radius of LLM training. We will discuss existing techniques and why they are undesirable, and the characteristics of an ideal technique for handling failures. 

\subsection{Distributed LLM Training Background}
\textbf{LLM Architecture.} An LLM is a sequence of transformer layers which are applied to an input sequence of tokens. Each transformer layer includes a set of operations, chief among which are attention operations and linear (matrix multiplication) operations.  

Distributed LLM training involves using multiple devices (often across several hosts) to accelerate training\cite{gpt3,llama3,gemini}. This is done 1) to leverage the compute capability of additional devices to reduce training time and 2) because the memory footprint of training is often several-fold the amount of HBM available on a single device, offloading to slower (e.g. host) memory is required if we do not distribute the memory footprint across enough devices. There are several techniques for parallelizing LLM training, and they vary in how effectively they achieve the two goals of distributed training; in practice these techniques are composed together. Below we describe the main techniques and their tradeoffs. 

\noindent \textbf{Data Parallelism.} Data parallelism (DP) was the first technique proposed for parallelizing DNN training\cite{pytorch_distributed}. In DP, the DNN parameters are replicated across multiple GPUs and the minibatch is partitioned across the replicas; each replica computes the gradients for its subset of the minibatch and then the replicas synchronize (allreduce) their local gradients. DP has two main downsides: 1) the DNN parameters + activations/gradients must fit entirely in within the memory of one GPU 2) DP can only partition a minibatch along the sample dimension. The first downside simply means that if a DNN is too large it cannot be trained with DP alone. The second downside is more complex and involves interaction with other parallelism techniques. Also, DP is limited by the training batch size and can not be increased arbitrarily increased since larger batch sizes degrades accuracy of the model.  \daiyaan{state here DP is limited to degree=minibatch size}
\ankit{perhaps also add a comment about ZeRO-1}

\begin{figure}
    \centering
    \includegraphics[width=.8\linewidth]{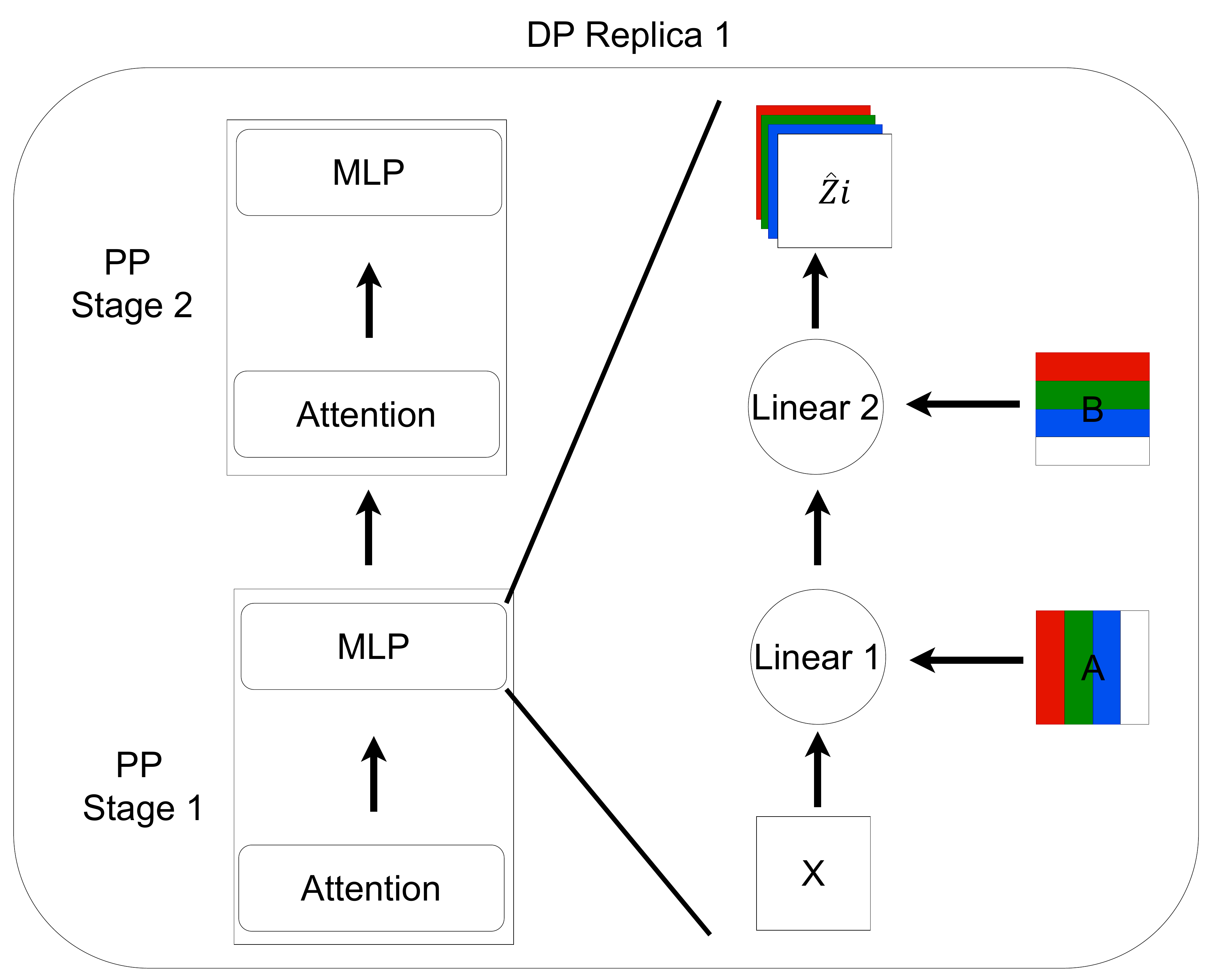}
    \caption{Hybrid parallelism: here we show one DP replica of a 2-layer transformer partitioned into two pipeline stages (left), we zoom in on one MLP layer to show how it is partitioned onto 4 GPUs using TP (right). Not shown are the TP partitions for Attention layers (which are partitioned across the same 4 GPUs as the MLP layer within a pipeline stage), see section \ref{sec:design} for a rigorous formulation of TP Attention and MLP.}
    \label{fig:mp-overview}
\end{figure}

\noindent \textbf{Pipeline Parallelism.} Pipeline parallelism (PP) aims to solve the memory constraints of DP by leveraging the layer-wise nature of DNNs\cite{pipedream,gpipe}. PP creates contiguous (ideally) uniform partitions of a DNN's layers (called pipeline stages), partitions the minibatch into several microbatches, and pipelines the forward and backward execution of the microbatches across the pipeline stages. In Figure \ref{fig:mp-overview} we show two transformer layers partitioned into two pipeline stages; for LLM training pipeline stage boundaries are often placed between transformer layer, though pipeline stages can contain more than 1 transformer layer. The major downside of PP is that the pipeline must be "flushed" between minibatches when the gradients are synchronized, creating periods called "pipeline bubbles" where GPUs are left idle. The pipeline bubble ratio (i.e. the proportion of time spent in pipeline bubbles versus active execution) is inversely proportional to the number of local microbatches per minibatch, and since DP can only partition along the sample/batch dimension using this makes using PP+DP problematic. Namely we face a severe consequence when scaling up a PP training job: as we scale up using DP (thereby reducing the local microbatches per minibatch) the pipeline bubble ratio increases as well. There are several works that attempt to reduce the bubble ratio, but they all depend on there being enough local microbatches per minibatch and therefore breakdown at higher DP scales. 

While PP does solve the issue of DP to an extent (by partitioning the model rather than the minibatch), PP is also limited here due to only being able to partition a DNN at layer boundaries (PP cannot partition an individual layer). This is problematic for two reasons: 1) as model sizes grow, typically the additional parameters are added through larger layers more than through larger number of layers 2) larger input sizes (e.g. longer sequence length) increase the execution time and memory-footprint of individual layers (which again PP cannot do anything about). For example, Llama-405B only has $1.6\times$ the number of layers as Llama-70B despite having $5.8\times$ the number of parameters. Moreover, despite attention compute growing quadratically with sequence length, Llama3.1 was trained on inputs up to 128K sequence length (up from 4K for Llama 2). 

\noindent \textbf{Tensor Parallelism.} Tensor parallelism (TP) parallelizes the DNN layers themselves across multiple GPUs. The application of TP varies depending on the type of layer (convolution, embedding, linear, attention, etc); TP for transformer models (mainly composed of linear and attention layers) was first introduce in \cite{megatron1} whereby TP partitions along the hidden/attention head dimension and operates with the parameters sharded (rather than replicated); thus TP solves the downsides of DP and PP. However, TP has its own downside: TP can induce very high communication volumes which are typically on the critical path of execution. To overcome this issue, TP is typically only used within \NVLdomain (where compute nodes are tightly coupled with very high bandwidth connection). In Figure \ref{fig:mp-overview} we show at a high-level how TP operates for the MLP layers within transformers: the parameters (A and B) are partitioned so that each GPU only computes a partial-sum ($\hat{Z}_i$) of the output. The high-volume communication induced by TP in this instance is to allreduce the partial sums in order to get the final output. For more details on TP for MLP and Attention layers, see Section \ref{sec:design}. 

\noindent \textbf{Hybrid Parallelism.} In practice, data, pipeline, and tensor parallelism are composed, a practice known as hybrid parallelism. The optimal degree of each type of parallelism is determined by the cluster architecture and the specific workload. The most common way this is done is as follows: 1) tensor-parallelism degree is mapped to the higher bandwidth scale-up domain; this is because although tensor-parallelism has the memory footprint advantage of being a partitioning technique and the scalability advantage of being independent of batch size, it induces high-volume blocking communications at every layer which quickly become the bottleneck without high-bandwidth interconnects between participants 2) pipeline-parallelism degree is set to the minimum required to fit the total memory footprint in HBM 3) data-parallelism degree is set to the number of replicas that can fit in the total available number of devices. Beyond DP, PP, and TP, there are other types of parallelism (e.g. context-parallelism\cite{ring_attention}, expert-parallelism\cite{switch_transformers}, fully-sharded data-parallelism\cite{zero}), but these can be seen as special versions of DP/TP. 

\subsection{Emerging Cluster Architectures}
\begin{figure}
    \centering
    \begin{subfigure}{0.46\columnwidth}
        \centering
        \includegraphics[width=1.15\columnwidth]{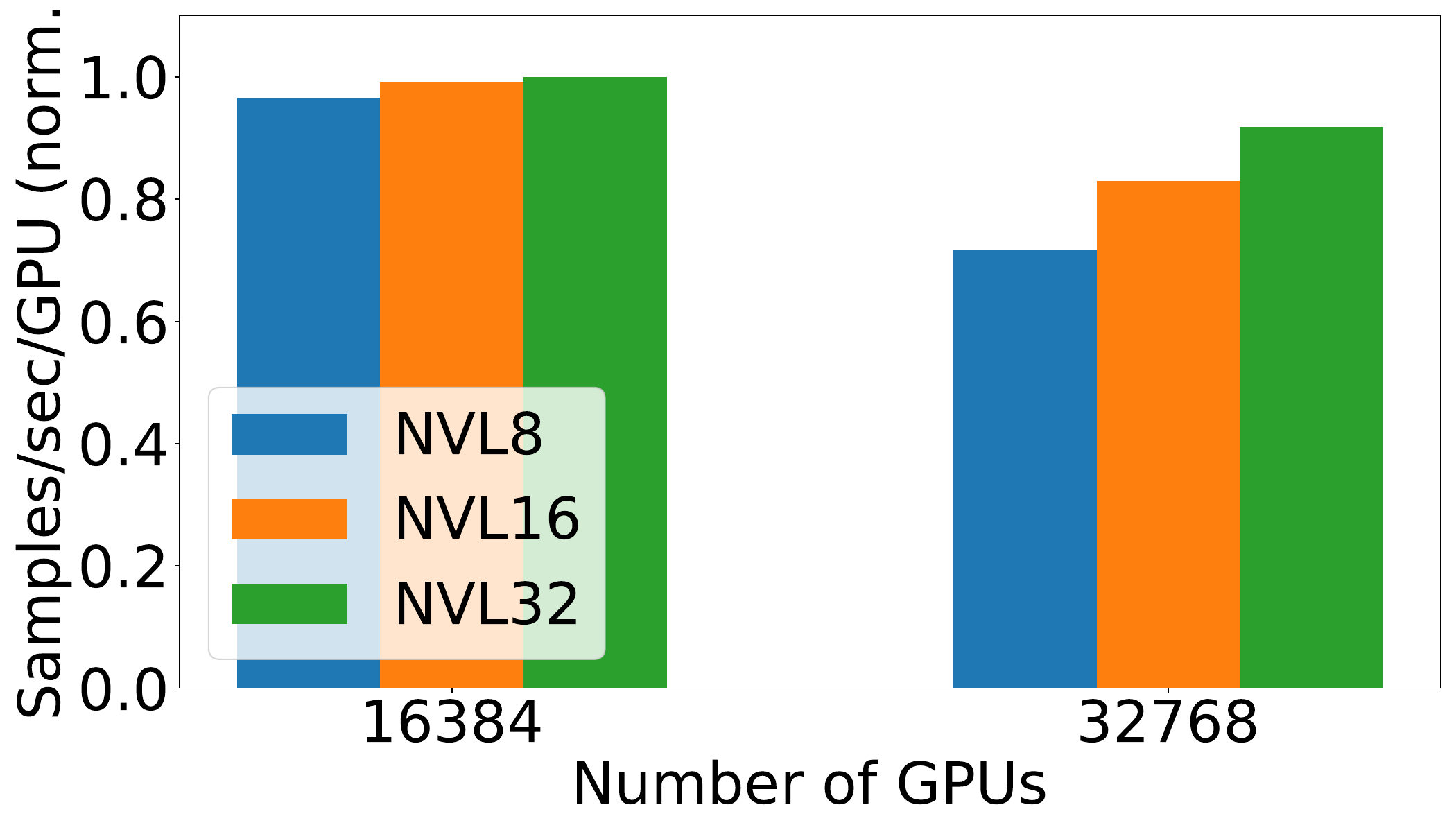}
        \caption{}
        \label{fig:nvl}
    \end{subfigure}
    \hfill
    \begin{subfigure}{0.46\columnwidth}
        \centering
        \includegraphics[width=1.15\columnwidth]{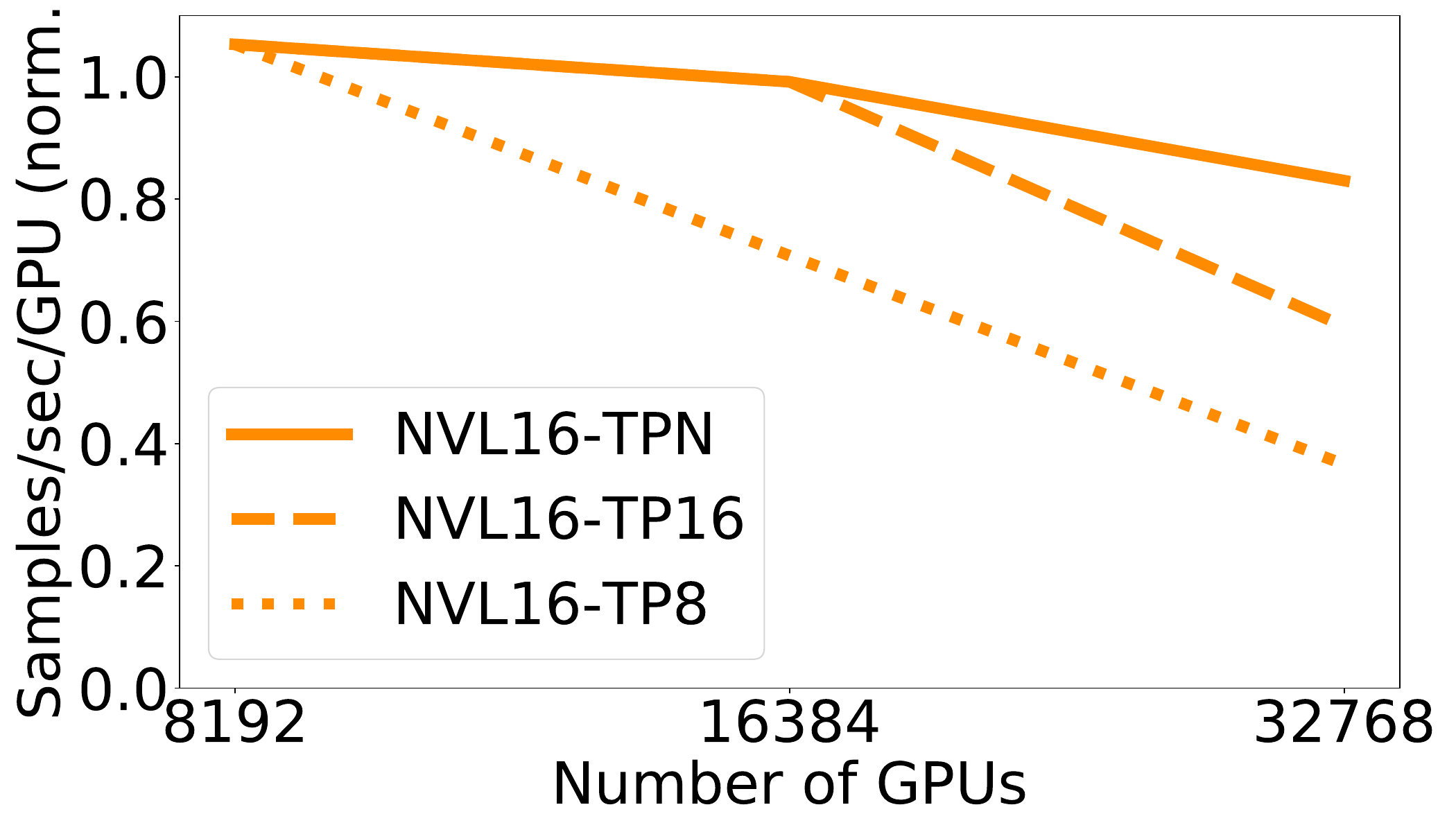}
        \caption{}
        \label{fig:tp}
    \end{subfigure}
    \caption{Effects of NVL domain size and TP degree on per-GPU throughput when scaling training. Throughputs are normalized to NVL32 at 16K GPUs. \protect\daiyaan{y-axis start at 0, fix the colors, make y-axis lost-throughput, make it clear that NVL16 is using unlimited TP}}
    \label{fig:motiv}
\end{figure}

Regardless of the specific hybrid-parallel configuration used, most LLM training jobs become increasingly communication-bound as they are scaled to run on many devices. To address this, hardware vendors are focusing on increasing the network bandwidth between devices. In particular, increasing the \NVLdomain such as TPU pods with upto 2K TPU processors\cite{tpuv4}, NVIDIA GB200 platform supporting a NVLink~(NVL) domain of up to 72 interconnected GPUs\cite{gb200-nvl72}. This is particularly desirable since it can enable higher scaling of tensor-parallelism. In Figure \ref{fig:nvl}, we train a 480B parameter LLM on clusters with varying NVL domain size (8K sequence length, 16M tokens/minibatch); at smaller scales (8K GPUs) higher NVL domain size does not significantly improve performance as we are not as communication-bound. As we scale up, a larger NVL domain size ensures we do not become communication-bound: at 32K GPUs we see a nearly 20\% difference in per-GPU utilization between NVL32 (87\%) and NVL8 (68\%).

\daiyaan{this is strangely worded, what we want to say is large TP is required to maximize throughput} With larger \NVLdomain sizes it may not always be the case that setting TP equal to domain size is the ideal hybrid-parallel configuration. Nonetheless, increasing TP is still a crucial part of maximizing throughput in large \NVLdomain clusters. In Figure \ref{fig:tp} we show a case of the same training job and cluster size with fixed NVLink domain size of 16; we exhaustively search the space of hybrid-parallel configurations and plot the optimal configurations' per-GPU throughput with TP degree limited to 8, 16, and unlimited. As training is scaled, we need higher TP degrees to maintain per-GPU throughput; this is because at higher-scales increasing DP/PP further will increase bubble ratio and increasing DP further will increase exposed all-reduce time. \daiyaan{should we show dlsim latency breakdown for this? at least comparing best TP16 config vs best TP8 config}
\dheevatsa{Furthermore, with the increased cadence of introduction of new hardware, its likely the a data-center will house multiple different platform possibly across multiple generations. Which requires large scale training jobs to deal with hardware heterogeneity, since these can be significantly different in terms of scale-up domain size, memory capacity, flops etc.}

\ahmet{should we explicitly mention unlimited TP configs in the fig 1b? assuming each config (8K, 16K, 32K) would have different TP}

% \begin{figure}
%     \centering
%     \includegraphics[width=\linewidth]{plots/motiv/nvl.pdf}
%     \caption{Caption}
%     \label{fig:nvl}
% \end{figure}

\subsection{Impact of Failures on Large scale-up domain Clusters}
\begin{figure}
    \centering
    \includegraphics[width=.9\linewidth]{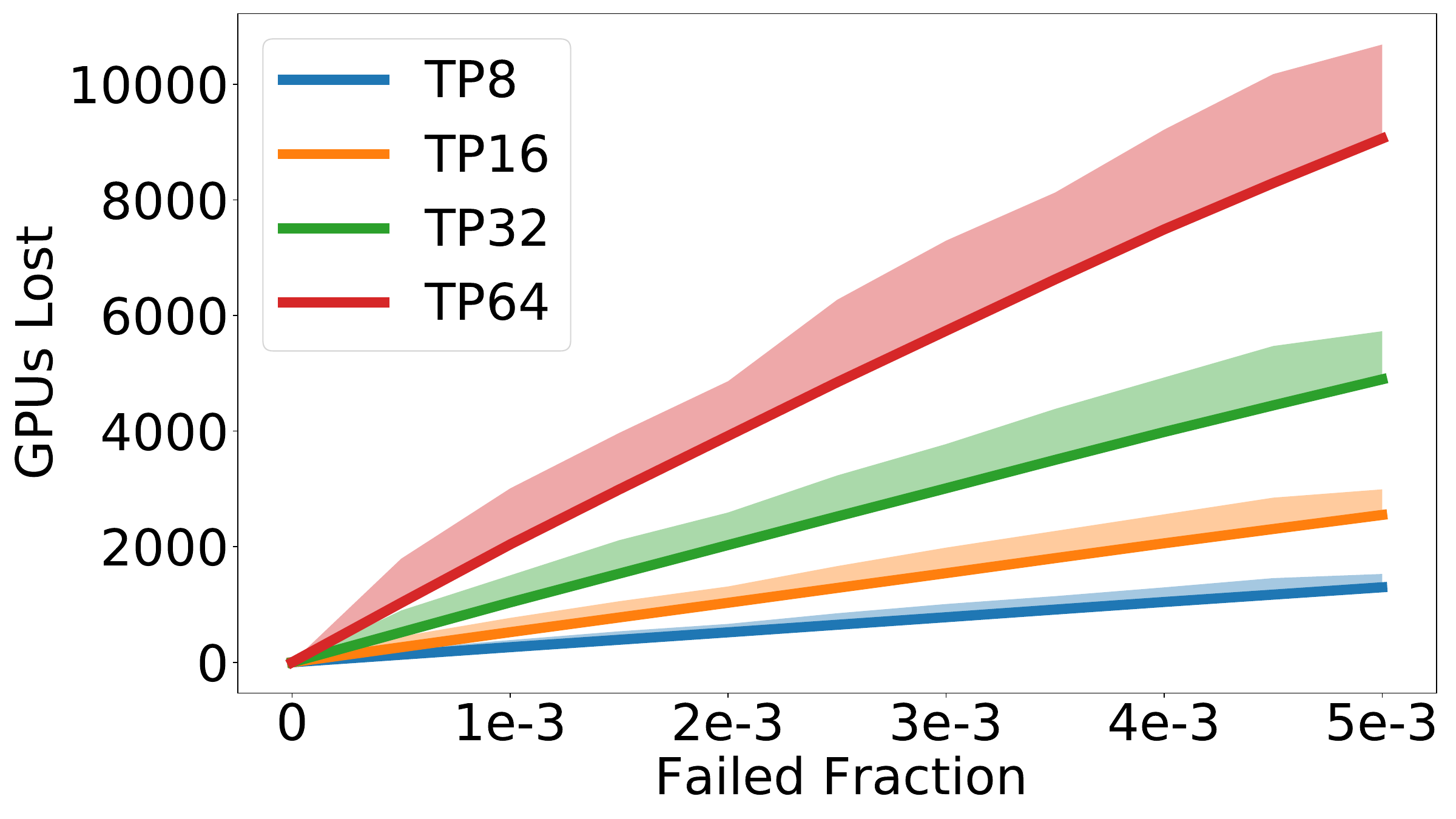}
    \caption{Larger TP/scale-up domains experience higher failure-amplification; the same number of failed GPUs impacts a larger fraction of the cluster. Solid lines are median lost and shaded are maximum.}
    \label{fig:tp-fail}
\end{figure}

We have shown how increasing TP is essential to scaling model pre-training on a large number of devices. 
We have also shown how large \NVLdomain reduces communication boundedness at large scales with increasing TP. Unfortunately, increasing the tightly coupled \NVLdomain size and TP degrees are double-edged sword: although they increase scalability they are also more vulnerable to failures. This is because a single GPU failure in a scale-up domain reduces the entire domain's ability to be used for TP. To illustrate the issue, we study a case with lager Nvlink domain GPU system (like the upcoming NVidia GB200, GB300 platforms). We plot the availability of a 32K GPU cluster as a function of the number of failed GPUs (uniformly distributed across the cluster) under different TP configurations in Figure \ref{fig:tp-fail}. Under the same number of failed GPUs, larger TP (requiring larger \NVLdomain) shows substantially lower availability due to individual failures being amplified by the \NVLdomain size. For TP64, as few as 0.1\% of GPUs being in a failed state is enough to reduce fleet availability to 94\%. 

\begin{figure}
    \centering
    \includegraphics[width=.85\linewidth]{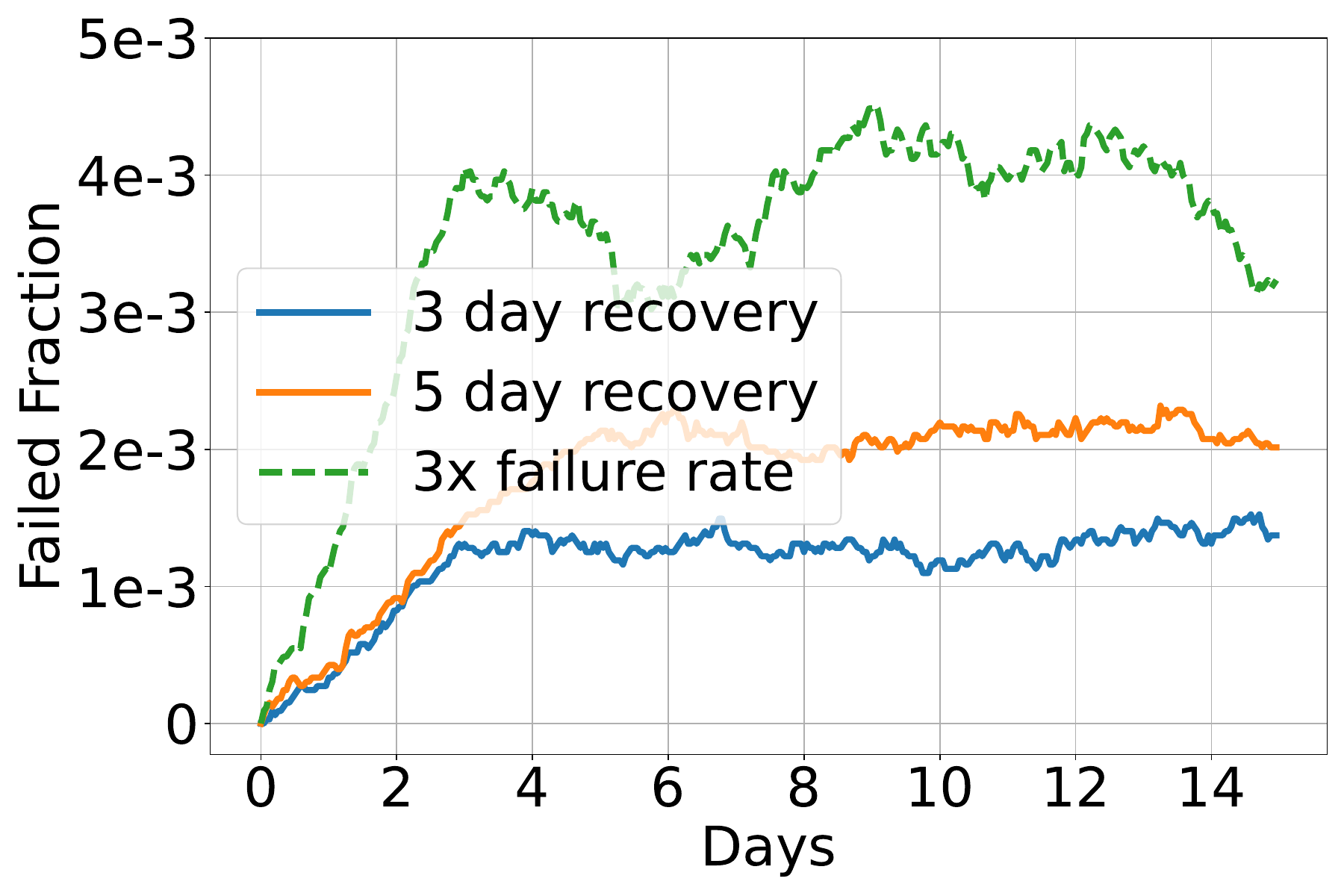}
    \caption{Observed and predicted failure rates and realistic failure recovery times result in high failure fractions.}
    \label{fig:fail-trace}
\end{figure}

Such failure scenarios are not uncommon either. In Figure \ref{fig:fail-trace}, we simulate a failure trace using the failure rate reported in Llama 3 training report\cite{llama3} for a cluster of NVIDIA H100 GPUs; we set 78\% of failures to hardware failures (as reported) with recovery times of 3/5 days (perhaps on the low-side for replacing high-demand hardware) and the remaining failures as software failures with recovery time 3 hours. In a 15-day trace, the cluster spends 81\% of time with > .1\% of GPUs failed, enough to drop TP64 availability to 94\%. 

With increasing complexity of the latest hardware (e.g. TPU-POD, GB200-NVL72) containing many more components (e.g. higher HBM capacity, larger number of cables for the higher scale-up bandwidth), it is reasonable to expect that the failure rates will be higher than those reported for NVL8 systems in the Llama~3 training report. Additionally, failure rates experience high variation over time and can experience major spikes: \cite{metafail} found failure rates in a cluster of 16K A100s can vary by $7\times$. Considering a hypothetical case with $3\times$ the failure rate compared to what was observed in the Llama report, we see ~$2\times$ higher peak concurrent failures, enough to drop availability to ~80\%. Therefore, it is clear that with the increasingly complex latest hardware, we need to design systems which are more resilient to failures.
\ankit{took a pass at simplifying this to justify why the failure rate is increased. }

%The failure rates observed in the Llama 3 training report were for a cluster of NVIDIA H100 GPUs; the latest hardware platforms are much more complex. For example, each NVIDIA Blackwell GB200 GPU has ~70\% higher TDP, has $2.4\times$ HBM capacity, and substantially higher cabling (due to a $9\times$ larger NVLink domain size and N$\times$ higher NVlink bandwidth). Moreover, the Llama3 training report was released approximately 2 years after the release of Hopper GPUs, and these systems tend to have much higher failure rates earlier in the release cycle. Thus, it is not unreasonable to assume that larger scale-up systems such as the TPU-POD,  GB200-NVL72 would be far more sensitive to failures and be able to deal with potentially higher failures due to sheer increase in complexity. For instance, considering a hypothetical case with $3\times$ the failure rate compared to what was observed in the Llama report, we see ~$2\times$ higher peak concurrent failures, enough to drop availability to  \daiyaan{how much?} 

\subsection{System objectives}
Existing solutions to the failure problem are less than ideal. The most straightforward solution to experiencing a GPU failure is to drop the entire DP replica to which the failed GPU belongs (we refer to this solution as \textbf{DP-DROP}). This solution causes even more failure amplification, and is also incompatible with stochastic gradient descent (SGD) (where the minibatch size must be fixed) due to it essentially dropping the samples that belong to the failed replica on every minibatch. The minibatch can be maintained if this technique is paired with spare DP replicas, but the use of spares is also less than ideal since it requires reserving a substantial amount of GPUs for an extended period of time which are essentially doing no useful work for most of that time. For example, TP64 in Figure \ref{fig:tp-fail} would require 2.5K spare GPUs (7.6\% of the 32K GPUs actually used for training) in order to maintain the required minibatch size at all times.

Instead, an ideal solution to the failure problem would:
\begin{enumerate}
    \item Require little to no spares to maintain the fixed minibatch size
    \item Experience a drop in throughput proportional to the percent of GPUs failed (i.e. no failure amplification due to NVL domain size / TP degree)
    \item Recuperate lost GPU utilization from failure amplification through GPU-sharing
\end{enumerate}
The only way to achieve goals 1 and 2 are to reduce failure amplification by enabling the use of partially failed NVL domains through TP reduction. However, this is not straightforward and requires solving several challenges:
\begin{enumerate}
    \item Reducing TP of one pipeline stage can bottleneck the other pipeline stages within the DP replica
    \item Reducing TP of one DP replica can bottleneck the other (healthy) DP replicas by delaying parameter synchronization 
    \item How to do parameter synchronization between two replicas using different TP is unknown, and the overhead induced by it could also slow training 
\end{enumerate}

NTP solves all of these problems. NTP uses nonuniform-tensor-parallelism, whereby some DP replicas/PP stages operate at reduced TP (due to partial failures) while healthy replicas/stages operate with the initial (higher) TP degree. Parameter synchronization is achieved with theoretically minimal overhead through careful shard placement and overlapped resharding. We also propose a new rack-design for large-NVL domains with improved electrical and thermal capabilities in order to sustain higher power-draw; combining this with NTP we can achieve uninterrupted training with virtually no throughput-loss with zero spares. 
\section{System Design}
\label{sec:design}

We present our system design for resilient TP. Our system contains a novel communication operation for synchronizing parameters between nonuniform TP replicas, and a dynamic power allocator for boosting throughput of unhealthy TP groups so they do not bottleneck healthy TP groups. Our system also contains a resource manager; for clusters with devices that do not have dynamic power features, our system minimizes the blast radius of failures and 1) regroups domains so unhealthy domains are in the same DP replica, reducing impact of subsequent failures and 2) makes any healthy devices that are impacted by a failure available for use by lower-priority jobs while the failures are repaired. Our system largely eliminates the need for spare devices, but still falls back on this technique when required. 

\subsection{Nonuniform Tensor Parallelism}
Tensor-parallelism for transformer layers are applied to the MLP blocks and self-attention blocks. An MLP block can be formulated as 
\begin{equation}
    Z=YB, \quad Y=GeLU(XA)
\end{equation}
where the input $X$ is multiplied by a parameter matrix $A$; the GeLU activation function is applied to the result of this $matmul$ to form the intermediate output $Y$ which is multiplied by parameter matrix $B$ to form the final output $Z$ of the MLP block. Tensor parallelism effectively parallelizes this sequence of operations by parallelizing each $matmul$. The output $Z$ is partitioned into partial-sums with $Y$ partitioned columnwise and $B$ partitioned rowwise.
\begin{equation}
    Z=\sum_{i} \hat{Z}_i=\sum_{i} Y_iB_i, \quad Y=[Y_1, ...Y_k], \quad B = \begin{bmatrix} B_1 \\ \vdots \\  B_k \end{bmatrix}
\end{equation}
Each tensor-parallel partition must thus first compute a column-partition of $Y$ which requires a full replica of $X$ and a row-partition of $A$.
\begin{equation}
    Y_i=GeLU(XA_i), \quad A=[A_1, ..., A_k]
\end{equation}
Typically, the columns/rows of $A$ and $B$ are partitioned into contiguous slices to create the tensor-parallel shards (see Figure \ref{fig:mp-overview}), but this is not strictly required. In fact, any column/row of A/B can be placed in any shard as long as the corresponding column/row from B/A is also placed in the same shard, i.e. it does not matter where each $\hat{Z}_i$ is computed as long as they are all summed (allreduced) at the end to form $Z$. Therefore, if one DP replica must operate at $TP=N_2$ due to failures while another replica can operate at $TP=N_1$ (with $N_1>N_2$), it is only a matter of partitioning (contiguously or non-contiguously) the columns/rows of A/B over $N_1$ shards for one replica and $N_2$ shards for the other. 

The challenge arises when we go to synchronize gradients (shaped and partitioned the same as A/B) across the two replicas; when $N_1=N_2$ and A/B are partitioned identically across the replicas, every shard only needs to synchronize with the single corresponding shard in the other replica(s) holding a copy of the exact same columns/rows of A/B (and the corresponding gradients). Naively sharding A/B across $N_1$ can induce small latency-bound communications that are unbalanced, resulting in not utilizing network bandwidth effectively and increasing its communication-boundedness. For example, if we contiguously partition A/B over both $N_1$ and $N_2$ (and $N_2 < N_1 < 2\times N_2$) then some shards will have to synchronize with 2 shards (in the other replica(s)) while some can synchronize with only 1. As $N_2$ gets closer in size to $N_1$, these two sub-shards will become less balanced with some sub-shards becoming so small that their synchronization will be highly susceptible to being latency-bound. For example, with hidden size 12K, $N_1=32$, $N_2=30$, in the TP30 group some shards will synchronize 375 columns/rows of A/B with one shard in the TP32 group and only 25 columns/rows with a different shard in the TP32 group. 
\ankit{could use a picture here to help understanding the concept of the sub-shard.} \daiyaan{does the numeric example here help clarify?}

\noindent \textbf{Shard-mapping algorithm} 
\ankit{@daiyaan may be change the wording to say redistribute the gradients as opposed to resharding. was a little confusing to me}
Ideally we would like to keep the 1-to-1 mapping between shards when synchronizing gradients (using $N_2$ GPUs in both replicas) since the bandwidth for synchronizing is limited by the reduced number of shards in partially failed replicas anyway. We would also like to keep the synchronization partitions contiguous so they can be fused as one operation to minimize latency overheads. This requires resharding gradients from $N_1$ shard groups to $N_2$ shard groups in the healthy replica. Since this resharding is done within TP groups (which are typically within an \NVLdomain) it can be done without bottlenecking the synchronization if overlapped, but we still want to minimize resharding time by maximizing parallelism leveraging higher bandwidth. 

We first assume that on unhealthy replicas, A/B (and their gradients) are sharded contiguously across $N_2$ GPUs. On healthy groups we pick $N_2$ GPUs out of the $N_1$ total GPUs on which the gradients of A/B will be contiguously sharded at the end of pre-synchronization resharding. Each $N_2$ shard must offload a number of columns/rows of A/B to the remaining $N_2-N_1$ GPUs in order to have a balanced sharding for computation; we enumerate all such rows/columns across all $N_2$ and iterate their placement across the $N_2-N_1$ offload GPUs. This ensures that every pairwise connection gets used to send an equal amount of data, fully utilizing the available bandwidth to do resharding. 

\noindent \textbf{Attention blocks} Attention blocks also contain two $matmuls$, with multi-head-attention applied to the output of the first $matmul$ before the second $matmul$. Attention blocks can be formulated as
\begin{equation}
    Z=\sum_{i}\hat{Z}_i=\sum_{i} MultiHead^{(i)}(X)W_O^{(i)}
\end{equation}
Each head is completely independent, computing
\begin{equation}
    MultiHead^{(i)}(X)=Attention(Q^{(i)}, K^{(i)}, V^{(i)})
\end{equation}
\begin{equation}
    Q^{(i)}=XW_Q^{(i)}, K^{(i)}=XW_K^{(i)}, V^{(i)}=XW_V^{(i)}
\end{equation}
Thus, analogous to the MLP blocks, attention blocks can be partitioned along the head dimension: as long as $W_Q^{(i)}$, $W_K^{(i)}$, $W_V^{(i)}$, $W_O^{(i)}$ are co-located we can compute $\hat{Z}_i$ on any GPU and sum them up to get the correct output. The partitioning dimension (k for MLP, heads for Attention) may not always divide evenly into the reduced TP size; in these cases there is some imbalance in the partition sizes. Since k is very large for MLPs, the imbalance is typically very small, though Attention usually has $O(10)$ heads creating potential for substantially more imbalance. Ultimately, these imbalance effects simply reduce the performance of the reduced TP replica. \daiyaan{this creates substantially coarser partitions we can shuffle using our algorithm, should maybe discuss it?} 

While NTP allows us to train with some DP replicas using reduced TP, the reduced TP replicas will have lower throughput; since all DP replicas must periodically synchronize parameters this can bottleneck the healthy replicas. To avoid this, DP replicas which have any partially failed scale-up domains train with reduced local batch size. This does reduce the minibatch size (by a very small amount); if this cannot be tolerated it can be mitigated through the use of spare DP replicas. The number of DP replicas impacted by failures can be minimized (and thus the number of spares required reduced) by actively packing partially failed NVL domains into as few DP replicas as possible. Alternatively, we also propose a redesigned rack which can boost the power supplied to partially failed scale-up domains so they can operate at the same throughput as (non-boosted) healthy domains. We discuss both solutions which can be used in tandem next. \dheevatsa{Additionally, non-uniform tensor parallelism enables potentially training on heterogeneous hardware, sizing up the TP dimension to optimize for the underlying HW and utilizing these for a single training job. In this work we don't focus on this aspect, hence drawing attention to capability for future works.}

\begin{algorithm}
\caption{Comp and Sync Rank Assignment}
\begin{algorithmic}[1]
\Require $k, n_1, n_2$
\State \textbf{Initialize:} \texttt{comp\_rank}, \texttt{sync\_rank} as arrays of size $k$
\For{$i \gets 0$ to $k-1$}
    \State \texttt{sync\_rank}[$i$] $\gets i // n_1$
\EndFor
\State \texttt{offload\_idx} $\gets 0$
\For{$i \gets 0$ to $n_1-1$}
    \For{$j \gets 0$ to $k // n_2$}
        \State \texttt{comp\_rank}[$j$] $\gets i$
    \EndFor
    \For{$j \gets k//n_2+1$ to $\text{length}(\texttt{sync\_rank}[i]) - 1$}
        \State \texttt{comp\_rank}[$j$] $\gets \text{range}(n_1, n_2)[\texttt{offload\_idx}+1]$
        \State \texttt{offload\_idx} $\gets (\texttt{offload\_idx} + 1) \mod (n_1 - n_2)$
    \EndFor
\EndFor
\end{algorithmic}
\label{algo}
\end{algorithm}

\subsection{Dynamic power allocation}
% We propose an alternative rack-design to enable power-boosting of partially-failed racks. Instead of dropping the local batch-size of partially-failed DP replicas, our design boosts the power of partially-failed racks so they can continue execution with the same (unreduced) local batch size at the same throughput as healthy racks. This, in combination with our NTP technique, enables training to continue at the same batch size and same throughput without any spare racks (in most cases). Our rack-design requires increased thermal and electrical capacities that we detail.

%Instead of sacrificing performance or relying on spare GPUs, our approach dynamically reallocates power within a rack, supplying up to 30\% additional power beyond the nominal TDP to the remaining functional GPUs within the rack. This compensates the performance loss caused by GPU failures, ensuring training throughput remains unaffected. 

%NTP consists of two key aspects: a) Algorithmic load balancing and b) Dynamic power allocation. 
When a subset of GPUs fail in a given model parallel partition, NTP redistributes the work to remaining GPUs, without decreasing the local batch size. Effectively, this increases the computations done by a given GPU. For example, when one of the GPU fails in a TP group with 8 GPUs, the remaining GPUs have to do 14.3\% more operations per GPU. With such a simple redistribution of work, this TP group of GPUs could become a bottleneck in a large scale synchronous training job. 

To alleviate the resulting performance degradation, we propose a flexible rack design that dynamically reallocates the power budget of the failed GPUs to other GPUs within the same rack, enabling them to maintain full throughput without reducing the local batch size. Our rack design can supply up to 30\% additional power beyond the nominal TDP to the remaining functional GPUs, allowing them to operate at a higher frequency and thus improving the performance/GPU. As a result, we are able to recover nearly all of the performance degradation due to GPU failures - without relying on spare GPUs in each rack.

In order to enable flexible power allocation within a rack, we provision the electrical and thermal components to meet the higher specifications. Specifically, the power delivery network at different levels must be designed considering the sum of the maximum power consumed by sub-components\cite{dynamo}. This includes the whips delivering power to the rack, circuit breakers, PDUs, PSUs and the voltage regulators delivering power to the GPUs. Note that this results in power oversubscription at a "row" level, i.e. the actual consumed power will be lower than the sum of maximum calculated above. 
%Note that such a design must also be aware of potentially higher transient requirements as well. 
Finally, the cooling capabilities of GPU and rack must now consider the additional power consumption (up to 30\% higher at GPU level). 

The above requirements are easily met in modern datacenters. In fact,  Nvidia GH200 GPUs have dynamic power balancing that allows GPUs to exceed the rated 700W budget and consume up to 900W (and 1000W in H200 variants \cite{mlperf-inference-nvidia}. This is an example where GPU originally designed for a particular power budget can operate at ~30\% higher power delivering additional performance.  Furthermore, the power budget of Nvidia GPUs have been increasing more than 50\% from Ampere (A100-SXM, 400W) to Hopper (H100-SXM, 700W) and Blackwell (B200-SXM/GB200, 1000W/1200W).  It shows that while cooling the GPUs is a hard problem, innovations in both air-cooling and liquid-cooling are allowing the chip makers to push the TDP to a higher limit in each generation. At the higher end, Cerebras \cite{cerebras} has shown the ability to cool an entire wafer consuming >~ 10kW. Therefore, we believe that the added electrical and thermal requirements do not pose an unreasonable challenge in our proposed flexible rack design.

By coupling these improved electrical and thermal solutions with the computational flexibility provided by NTP, our design maintains stable training throughput during partial failure scenarios without the overhead of redundant hardware. This demonstrates the practical feasibility and significant advantages offered by dynamic power allocation in datacenter systems.

\daiyaan{@bhargava @ahmet}

\subsection{Resource manager}
If a training job is using PP, and a DP replica contains some partially failed PP stages, the remaining healthy PP stages will be bottlenecked by the partially failed stages. One way to mitigate this is through PP stage re-balancing, but this is a very complex technique which may not be compatible with more complex PP schedules (e.g. 1F1B) and would also induce highly complex (i.e. many-to-many) PP stage parameter synchronization (since different DP replicas would be operating with different PP stage partitions). Instead, we opt to "pack" partially failed scale-up domains into as few DP replicas as possible, and have any DP replica containing any partially failed scale-up domains operate at reduced domain / TP size. When a failure occurs, the job must be restarted anyway; when the job is restarted the process-group ranks are assigned so that unhealthy racks are packed together by being placed in the lowest ranks. This minimizes the number of DP replicas affected by failures, thus minimizing (though not eliminating) the PP stage bottleneck issue. The remaining healthy + bottlenecked scale-up domains are forced to operate at a lower domain / TP size than they potentially could, but the leftover idle GPUs from these healthy domains can be made available to run other workloads rather than remain idle. 
\section{Implementation}
\subsection{NTP Implementation}
\label{ntp-impl}
\begin{figure}
    \centering
    \includegraphics[width=1.05\linewidth]{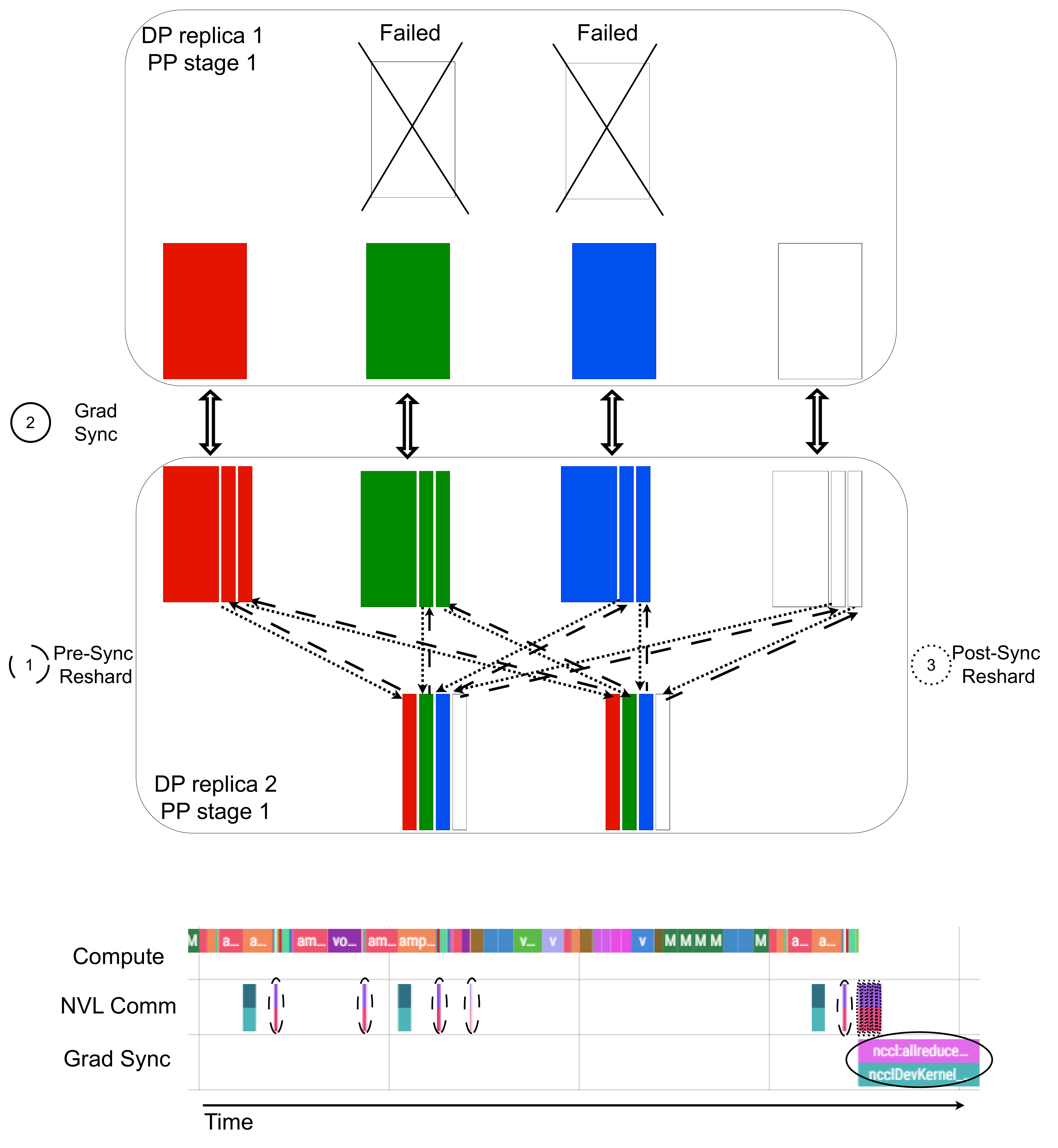}
    \caption{Illustration of NTP (top) with a PyTorch trace of our prototype (bottom). Presync-reshard (dashed) runs in the NVL comm stream and is overlapped with computation (which is completely unaffected and is identical to healthy operation), postsync-reshard (dotted) runs in the NVL comm stream and is overlapped with gradient allreduce (solid) which incurs slightly increased volume. The trace shows 5 different gradients being resharded as soon as they are computed, all 5 gradients being synchronized together in one allreduce operation, and then being resharded post-sync one after another.}
    \label{fig:ntp_sync}
\end{figure}

Figure \ref{fig:ntp_sync} (top) shows a high-level overview of how NTP resharding is overlapped with gradient synchronization. We build our implementation of NTP on top of NVIDIA Megatron framework\dheevatsa{needs reference}\cite{megatron-framework}. The resharding before allreduce is implemented as part of a PyTorch backward hook (i.e. it is overlapped with the final backward pass); the hook is intended for marking a gradient as ready-for-synchronization (when all gradients is a bucket are marked ready the whole bucket is synchronized), our implementation reshards a gradient before marking it as ready. The post-synchronization resharding is performed with the last bucket's allreduce; this is because Megatron uses {\small CUDA\_DEVICE\_MAX\_CONNECTIONS=1} for performance stability/predictability, and to ensure that prior gradients are finished synchronizing before post-sync reshard we wait until the final bucket all-reduce. In our evaluations, we breakdown our implementation's overhead as 1) pre-sync reshard's overhead on the final backward pass 2) all-reduce overhead from increased all-reduce volume 3) post-sync reshard overhead on exposed all-reduce time. In Figure \ref{fig:ntp_sync} (bottom) we show a PyTorch trace of our implementation showing the pre-sync reshard ops (running on the same stream as TP comm ops) overlapped with the backward pass and the post-sync reshard ops overlapped with the final allreduce bucket. \daiyaan{should we add a figure showing pytorch trace?}\dheevatsa{Yes - that would help}

% \begin{figure}
%     \centering
%     \includegraphics[width=\linewidth]{figures/trace_small.PNG}
%     \caption{Caption}
%     \label{fig:trace}
% \end{figure}

\noindent \textbf{Pipeline-parallel communication} In transformer layers, the output of TP MLP/attention blocks are replicated across TP shards. Since PP stage boundaries are always between transformer layers, the output of PP stages are also always replicated across TP shards. Activations/gradients are sent between stages using an optimization from \cite{megatron2}; this method for sending activations between pipeline stages is bound by the cross-stage aggregate infiniband/ethernet bandwidth. The P2P communication induced by PP is a very small part of end-to-end latency. \daiyaan{should we show dlsim breakdown for this?} Nonetheless, if one PP stage has a reduced TP size it proportionally reduces the aggregate cross-stage bandwidth. For NTP (without power redistribution) all stages operate at reduced TP, so simply sending the activations (with lower bandwidth) is enough. For NTP-PW, where reduced TP stages may have to send/receive activations to healthy TP stages, activations are exchanged at the reduced TP size with the proportionally lower bandwidth and then (if needed) they are broadcasted to the additional GPUs in the larger TP group; the broadcast occurs within an \NVLdomain and can be overlapped with receiving the activations, effectively adding zero overhead. In our simulations for NTP and NTP-PW, we add the overhead of sending/receiving activations with lower cross-stage bandwidth and treat any needed broadcast as fully-overlapped. 

\subsection{Performance modeling and estimation}
Although we evaluate and show a proof-of-concept design on current systems, the primary use case for NTP is systems with larger \NVLdomain such as the B200-NVL72 systems.
Due to the lack of wide availability of such systems, we evaluate the benefits of NTP using a high-level performance simulator that models the performance of large-scale multi-node GPU systems. This simulator is fairly sophisticated, with detailed modeling of the underlying GPU micro-architecture, LLM application parallel mapping as well as communication/network behavior. Since the amount of work performed per GPU in an LLM is fairly the same, the simulator partitions the work on a per GPU basis based on the parallelism configuration being used. In the results section~\ref{subsec:sim_val} we show correlation studies comparing the simulator's projected performance to measurements on actual system, thereby establishing the fidelity of the simulator and hence providing confidence/backing for the projected numbers. 

Specifically, the LLM is defined as a graph which is partitioned based on the parallelism strategy with the associated communication operations inserted in the graph. We then model the performance of the different compute and communication operations happening on a single GPU taking into account the GPU micro-architecture and the scale of the system. 
In addition, the simulator also allows us to model different computation and communication overlap strategies. 
As such, the simulator is capable of scaling to both large \NVLdomain such as NVL72 (or larger), scaled-out to larger clusters.
Besides estimating the performance at scale of the application, the simulator also estimates its power consumption and boosts the performance of the device similar to a power management system. We use this to evaluate NTP with dynamic power redistribution.

\daiyaan{simulator assumptions when scaling}
\ankit{@daiyaan, done with this section unless we need more details}
\section{Experimental setup}

\subsection{Prototype evaluation}
\label{subsec:poc-desc}
We build a prototype of our system as a proof-of-concept and for some sensitivity studies for NTP overhead. We evaluate our prototype on 2$\times$ DGX-A100 machines. Each machine has 8$\times$ 80GB-A100 GPUs with 600GBps NVLink bandwidth, 8$\times$ 200Gbps HCAs connecting to a shared infiniband network. We profile training of LLMs with hidden size 12288 and 6144, attention head dimension 128, FFN dimension equal to 4$\times$ hidden size, sequence length between 4K and 16K. We profile training of 2 DP replicas (one on each machine) of 1 PP stage (i.e. no PP) to isolate and measure the DP+TP overheads of our method (since the PP overheads are negligible); each DP replica trains 1-3 layers on a single local batch with 1-2 samples before parameter synchronization. We vary the TP size of each replica using NTP. 

\subsection{Validating our performance model}
We validate our performance model's accuracy at large-scales and at different TDPs. For these experiments we profile and simulate training on DGX-H100. We train a variety of workloads described in \ref{subsec:sim_val}.

\subsection{Large-scale simulations and sensitivity studies}
To evaluate NTP and NTP-PW at large-scale and large NVL domain size, we use our simulator. We simulate training on a cluster with 32K B200 GPUs with 189GB memory, NVL domain size of 32 GPUs (1.8TBps per GPU), 800Gbps infiniband per GPU. %, baseline of 1000W/GPU TDP.
We train a 480B parameter model with hidden size 20480, 128 attention heads, FFN dimension equal to 4$\times$ hidden size, and 100 layers. We train the model with 16K sequence length, 16M tokens / minibatch. 

As mentioned in section \ref{sec:design}, DP replicas which contain any partially failed replicas must either reduce their local batch size or power boost the partially failed NVL domains to avoid bottle-necking healthy DP replicas. Our simulation workload operates at TP32, and we support TP reduction to TP30 and TP28; for the reduced TP sizes we use our simulator to find the maximum local batch size (for non-power-boosted) and minimum operating power (for power-boosted) for the iteration time on the reduced TP replicas to be less than or equal to the iteration time of the healthy replicas. We list all the configurations and their performance in Table \ref{tab:setup}.

\begin{table}
    \centering
    \renewcommand{\arraystretch}{1.2}  % Adjust row height for readability
    \begin{tabular}{|l|c|c|c|}
        \hline
         & \textbf{Local bs} & \textbf{Power}  & \textbf{Rel iter time}\\ \hline
        \textbf{TP32} & 8 & $1\times$ & 1 / .994 \\ \hline
        \textbf{TP30} & 7 & $1\times$ & 1.002 \\ \hline
        \textbf{TP30-PW} & 8 & $1.15\times$ & .978 \\ \hline
        \textbf{TP28} & 6 & $1\times$ & 1.003 \\ \hline
        \textbf{TP28-PW} & 8 & $1.3\times$ & .999 \\ \hline
    \end{tabular}
    \caption{Simulated training settings \protect\daiyaan{relative power values}}
    \label{tab:setup}
\end{table}
\section{Evaluation}

\subsection{Main results}
\label{subsec:main-result}
\begin{figure}[t]
    \centering
    \includegraphics[width=\linewidth]{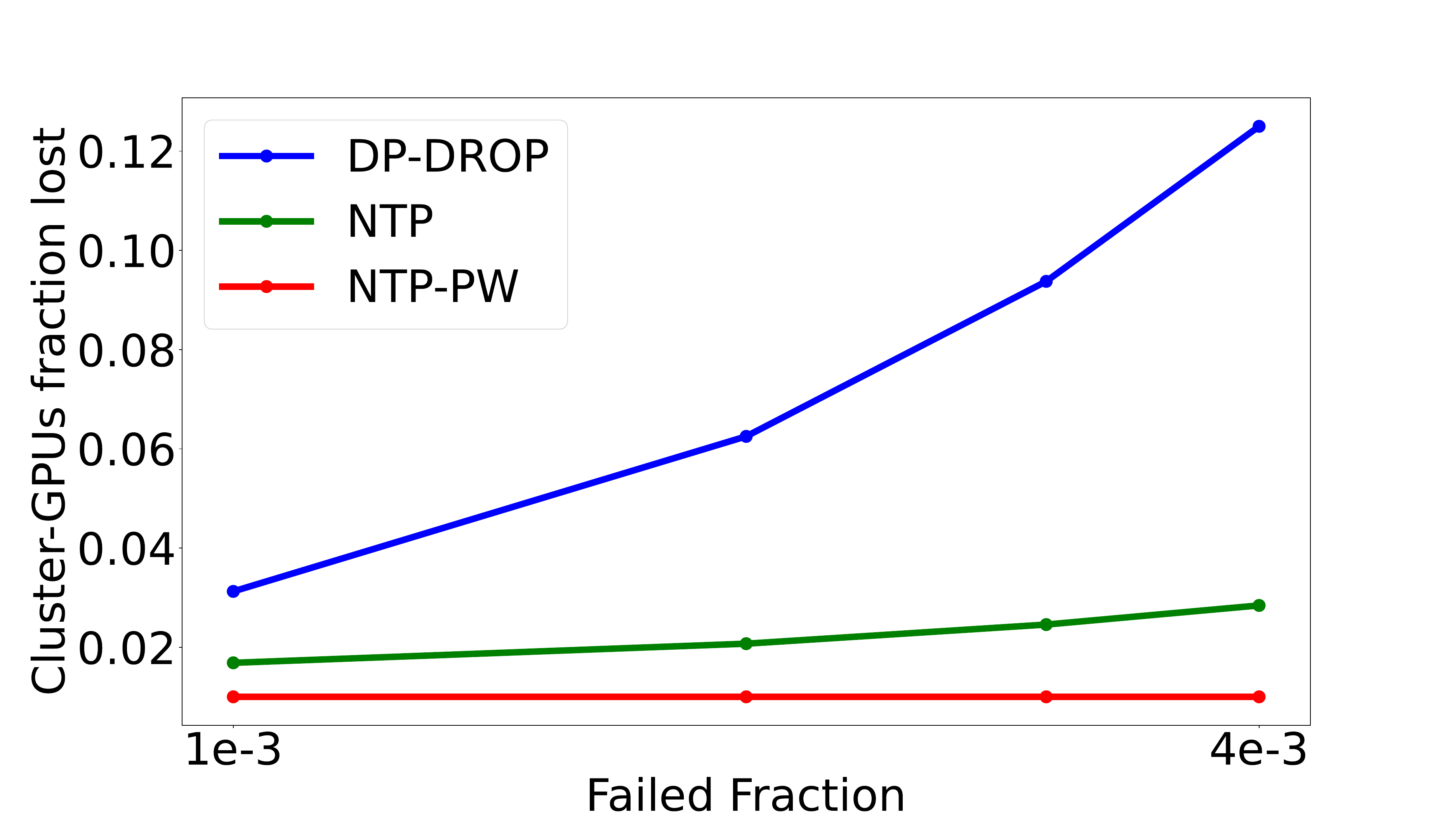}
    \caption{NTP vs DP-DROP total GPUs lost as a function of the fraction of GPUs down. NTP and NTP-PW reduce failure amplification compared to DP-DROP.}
    \label{fig:main-throughput}
\end{figure}

In Figure \ref{fig:main-throughput} we vary the fraction of GPUs failed on the x-axis according to the failed fractions observed in Figure \ref{fig:fail-trace}; for each failed fraction, we calculate the throughput-loss for each fault-tolerance method. Since the distribution of the failed GPUs can vary the resulting throughput, we sample a large number of failure scenarios for each fraction, and plot the mean across the scenarios. For DP-DROP, we see up to a 12\% drop in throughput (with a proportional drop in minibatch size). NTP reduces the drop to 3\%, and NTP-PW maintains <1\% throughput with up to 4e-3 failed fraction. 
\ankit{@daiyaan, needs description of DP-DROP, this is the first time we use it}

\begin{figure}[t]
    \centering
    \includegraphics[width=.9\linewidth]{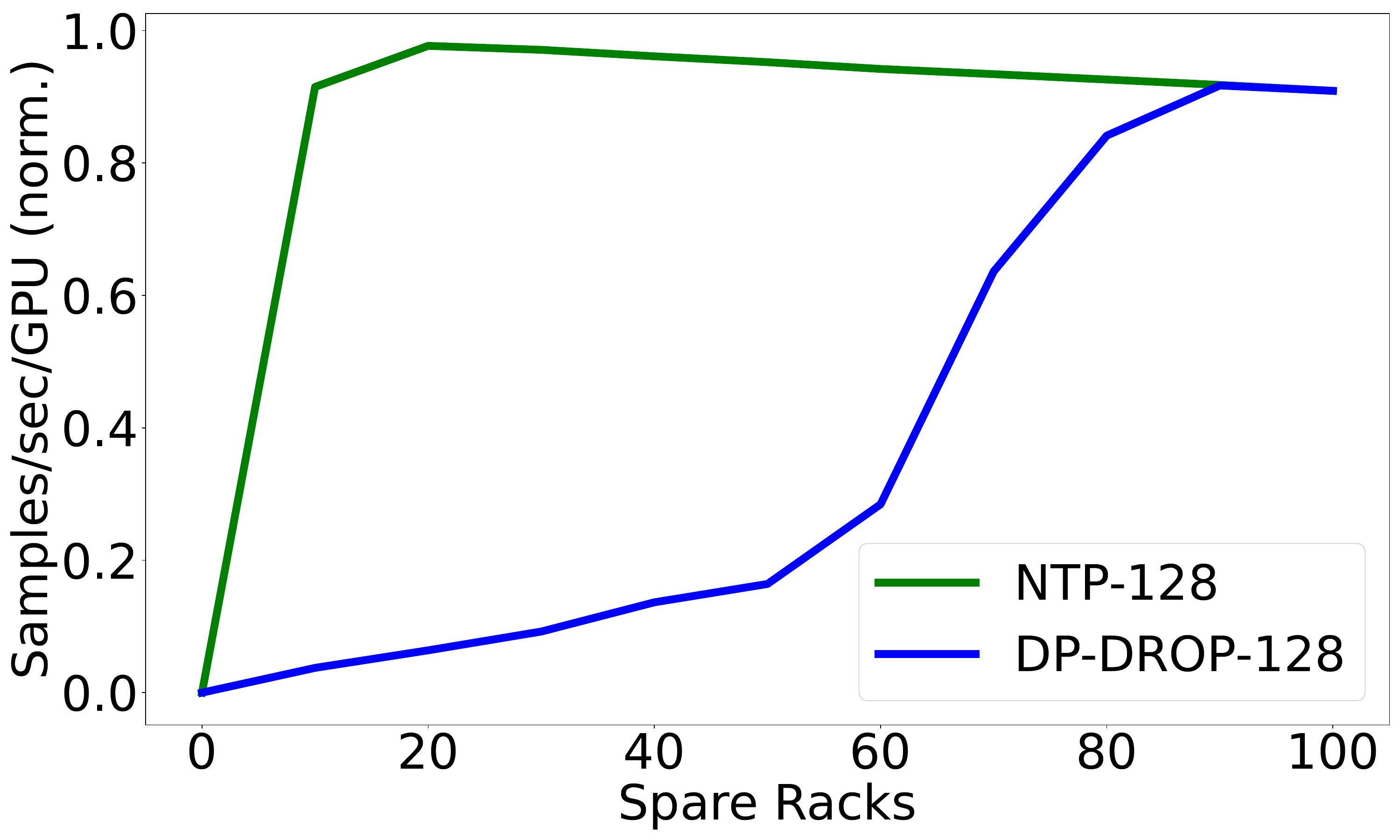}
    \caption{DP-DROP requires 90 spare racks to maintain full mini-batch size. NTP only requires 2 spare DP replicas (in this case 16 racks) to maintain full mini-batch size under this failure scenario, NTP-PW is able to maintain it with zero spares. Throughputs are normalized to a cluster with no failures. \protect\daiyaan{point back to steady-state failure rate from earlier}}
    \label{fig:main-util}
\end{figure}

Conversely, in Figure \ref{fig:main-util} we treat the minibatch size as fixed; when the number of failures reduces the minibatch size below the target size, training is paused until enough failures recover to satisfy the minibatch size. We use the same failure rate as observed in Llama3.1 with a hardware failure recovery time of 5 days. We increase the number of spare NVL domains, and plot the throughput-per-GPU. NTP requires only 16 spare NVL domains to be able to maintain training with no pausing; this is because this training job uses 8 NVL domains per DP replica, and NTP (with no spares) never experiences a throughput drop larger than the equivalent of dropping two DP replica. DP-DROP requires 90 spare NVL domains for uninterrupted training, and this further reduces throughput-per-GPU since it requires 90 extra NVL domains for the same throughput as NTP with 16 spares. NTP-PW requires zero spares under this failure scenario for uninterrupted training and <1\% throughput reduction compared to experiencing no failures.

\subsection{Prototype Evaluation}
\begin{figure}[]
    \centering
    \includegraphics[width=\linewidth]{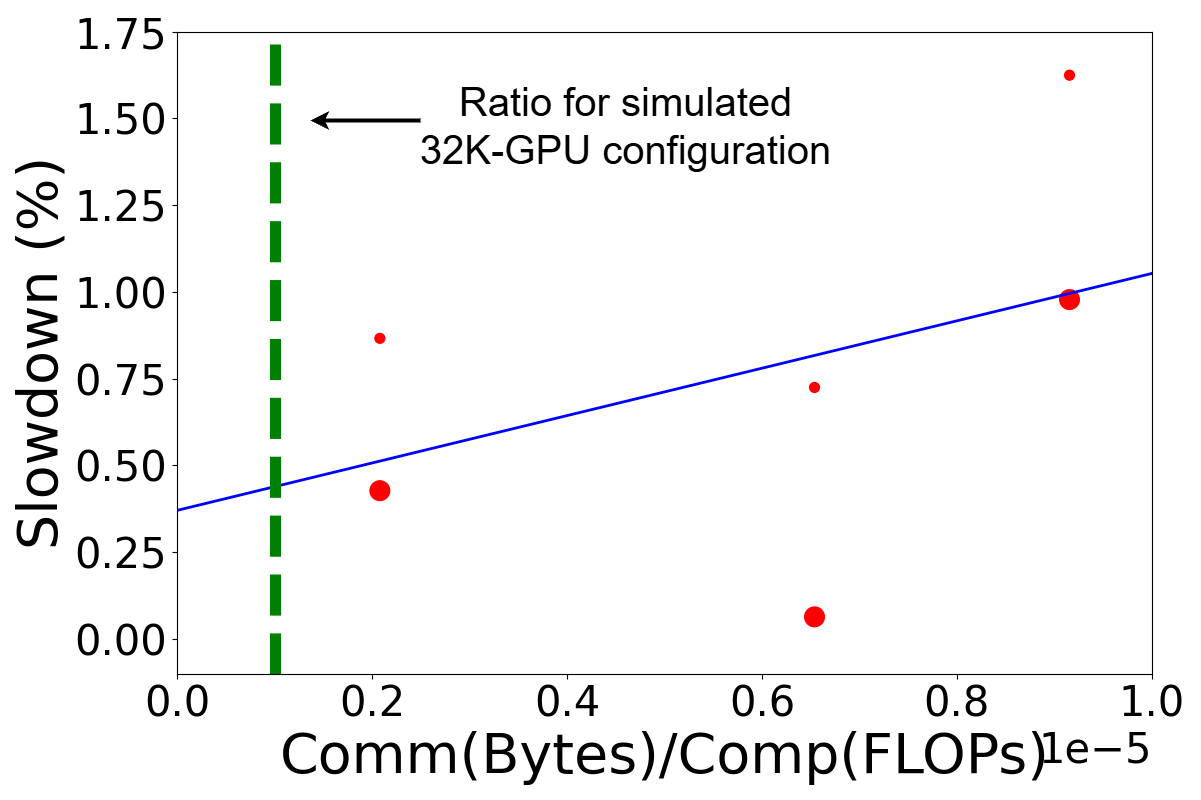}
    \caption{We find a strong relationship between the communication:computation ratio and the final backward pass slowdown from NTP. Small dots are for TP reduction of >50\% and large dots are for TP reduction < 50\%. Marked is the ratio for the workload we simulate in \ref{subsec:main-result} and \ref{subsec:sens}: large model size, large TP size, and small TP reduction put us comfortably in the <1\% slowdown regime. \protect\daiyaan{assert that we are always on the left side of this graph}}
    \label{fig:bwd-slowdown}
\end{figure}

We evaluate our prototype described in \ref{subsec:poc-desc} to measure the overhead of NTP. Since the pre-sync reshard is overlapped with the final backward pass, we measure its overhead on the backward pass. In Figure \ref{fig:bwd-slowdown}, we profile a variety of workloads with different hidden size and sequence length (described in \ref{subsec:poc-desc}). We train with one DP replica with TP8 and local batch size 2 and the other replica at reduced TP with local batch size 1, and measure the backward pass slowdown (compared to training with TP8 and local batch size 2 on both replicas) on the TP8 replica (which must do resharding). We hypothesize that the slowdown is a function of two values: 1) the total backward computation, because the larger the backward computation the more opportunity there is for the resharding to be overlapped 2) the maximum number of bytes sent/received by a GPU for resharding, since this has a direct effect on how long resharding takes. (1) is affected by the model size and sequence length and (2) is affected by the model size and the reduced TP degree (the smaller the reduced TP degree the higher the communication induced by resharding). We divided (2) by (1) to calculate a communication-to-computation ratio and plot this on the x-axis of Figure \ref{fig:bwd-slowdown} and the backward pass slowdown on the y-axis. The first observation is that there is a strong linear relationship between the communication-to-computation ratio and the slowdown, confirming our hypothesis. The consequence of this is that the larger the model or the longer the sequence length the smaller the slowdown will be because computation grows quadratically with both while reshard cost grows linearly only with model size and does not grow with sequence length. The second observation is that smaller reduced TP sizes (i.e. larger fraction of initial TP size fails) have larger slowdown; this is because although our implementation overlaps the majority of the resharding with the backward pass there are still some exposed resharding operations; with a larger TP reduction (requiring more resharding volume) there is also an increased likelihood of some resharding operations being exposed. Nonetheless, all workloads and settings have at most 4\% slowdown for the final backward pass. Additionally, the communication-to-computation ratio for our simulation experiments is much lower than any of the prototype experiments (due to a much larger model size, much larger TP size, and relatively small TP reduction) putting us comfortably in the <1\% slowdown regime.

\begin{figure}
    \centering
    \includegraphics[width=\linewidth]{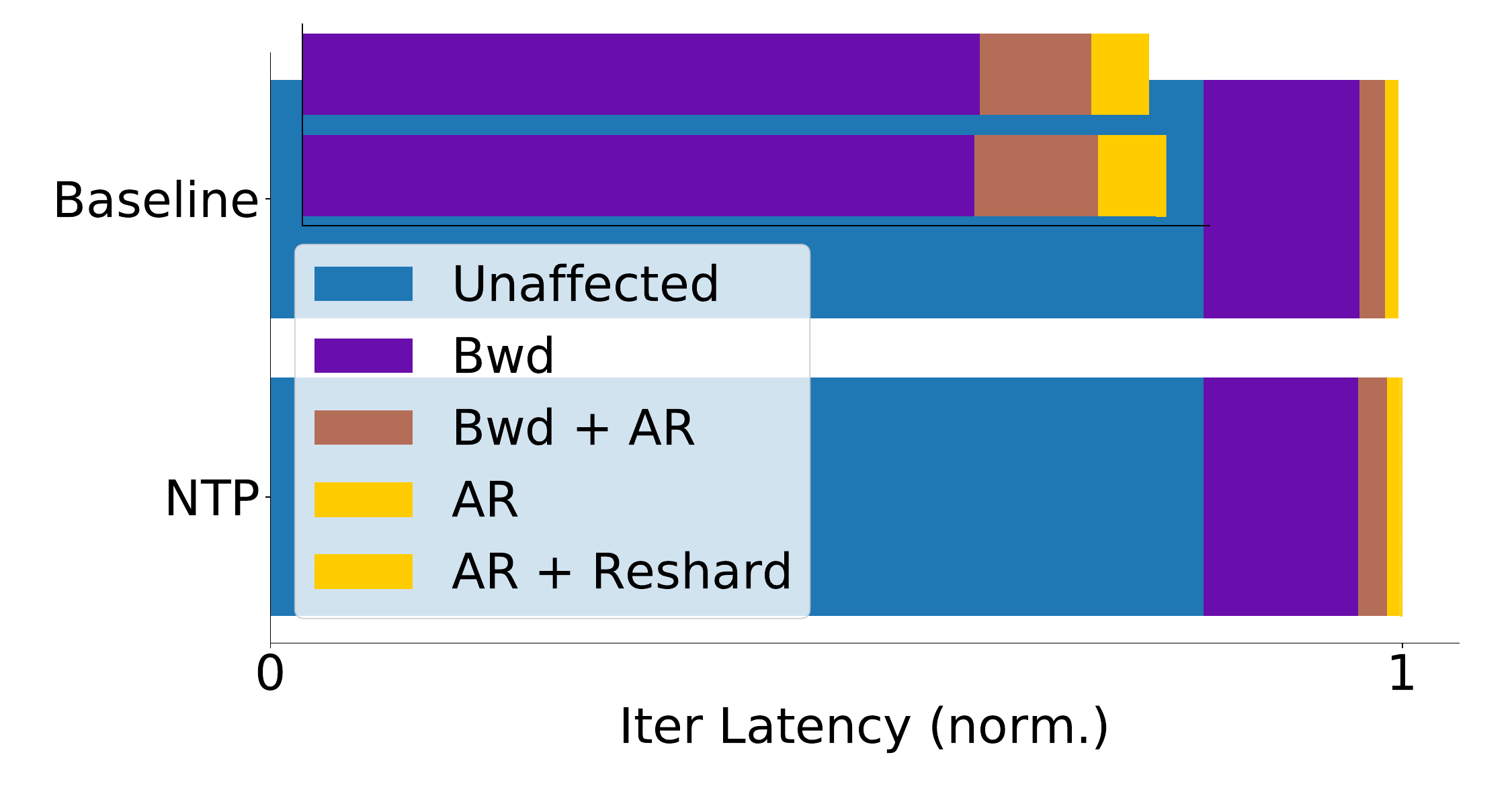}
    \caption{The majority of iteration time is unaffected by NTP (blue). We zoom in on the portion that does experience slowdown to show 1) the total backward compute slowdown from pre-sync reshard is limited (purple + brown) 2) there is some slowdown from the gradient allreduce volume increase (brown + yellow) 3) the two overheads have some amount of overlap so the total overlap is less than the sum (brown) 4) the post-sync reshard is totally overlapped with the allreduce (hatched yellow).}
    \label{fig:e2e}
\end{figure}

The (already very low) backward pass overhead induced by pre-sync resharding is only incurred by the final local batch's backward pass (since resharding + synchronization only occurs once all gradients for all samples are computed). In Figure \ref{fig:e2e} we show the end-to-end overhead breakdown of NTP for the model with 12K hidden size with 8K sequence length using reduced TP6. The majority of the iteration time is unaffected by NTP, with the overheads coming from the backward pass slowdown and the increased allreduce time. The allreduce time increases proportionally to the TP reduction, since a reduced number of GPUs must synchronize the same total number of gradients. However, since allreduce itself is also overlapped with the backward pass, both the extra backward time and extra allreduce time increases the amount of overlapping. Additionally, the post-sync reshard is completely overlapped with the allreduce. The result is an end-to-end slowdown of <1\%, the majority of which is caused by the increased allreduce time. 

\subsection{Simulation validation}
\label{subsec:sim_val}
In Figure \ref{fig:sim-scale} we train a variety of model sizes (8B-175B) with different sequence lengths (2K-8K) at many different scales (8-512 GPUs) with both FP8 and BF16 on DGX-H100 machines. We exhaustively search for the best parallelism configurations for each workload and plot the observed throughput on physical machines against the predicted throughput by our simulator. As shown, our simulator is highly correlated with observed performance. 

In Figure \ref{fig:sim-pow} we train two model sizes (15B and 340B) using DGX-H100 GPUs using a range of per-GPU power budgets; % [BG: No]} 
we plot the observed training performance against the predicted performance from our simulator. As shown, our simulator exhibits very high correlation against measured data. 

% \begin{itemize}
%     \item First we show that the simulation results are aligned with our small-scale physical cluster results. 

%     \item Comparison of our system with KSO (throughput and batch size reduction)

%     \item Ablation of different components of our system (fault-tolerant TP, batch size reduction, dynamic regrouping of healthy/unhealthy racks, side jobs on unused GPUs in partially unhealthy DP replicas, power redistribution)

%     \item Study on workload variation (model size, sequence length, minibatch size, parallelism mapping)

%     \item Different failure rate scenarios

%     \item different NVL/TP sizes, cluster scales

%     \item different simulation assumptions (bandwidth, compute, etc)

%     \item Some result on spare capacity required vs batch-size tolerance (how throughput constraints affect this)
% \end{itemize}

\begin{figure}[H]
    \centering
    \includegraphics[width=\linewidth]{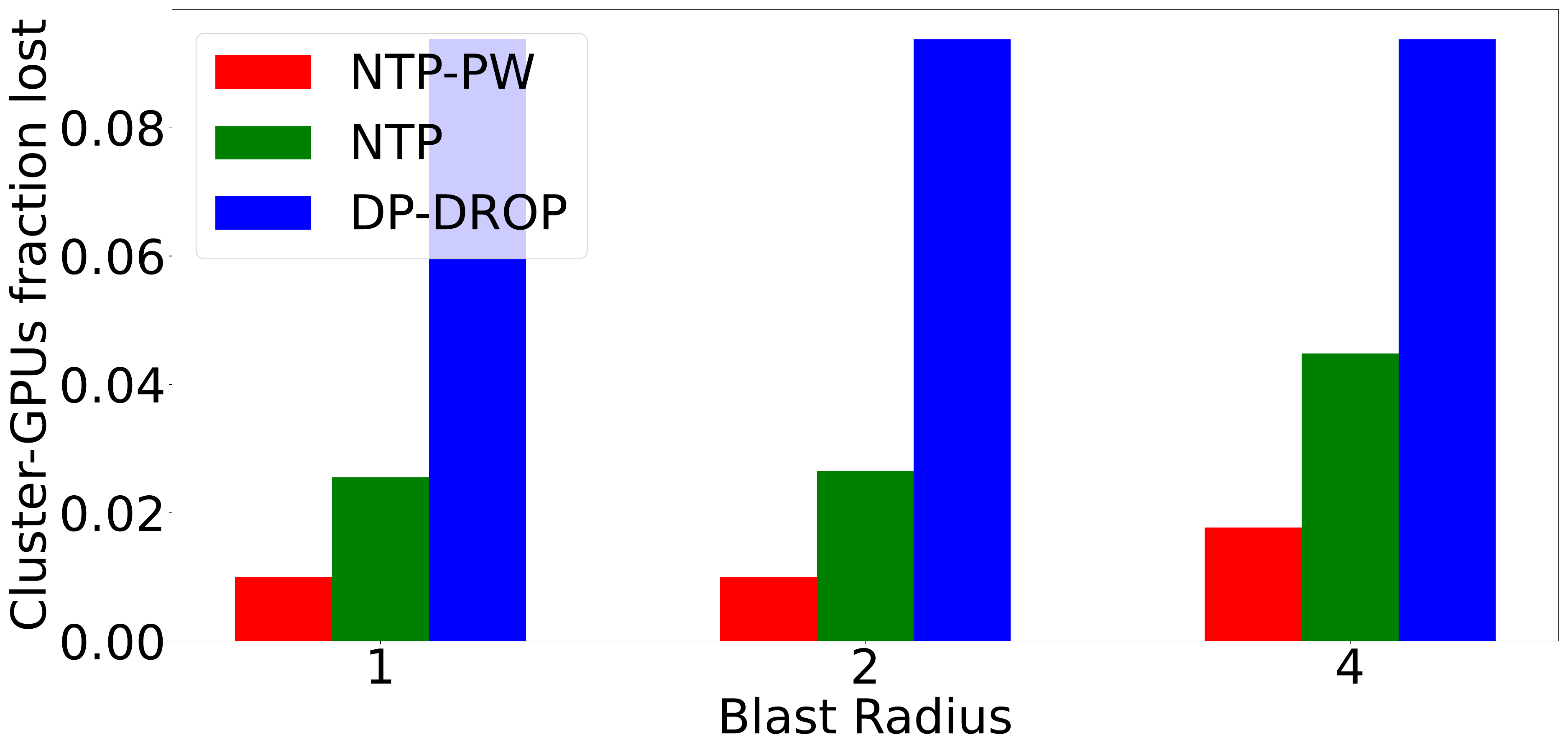}
    \caption{We plot the fraction of total cluster GPUs lost due to failures at various failure blast-radii. Larger blast-radii reduce NTP's performance, but it still outperforms DP-DROP.}
    \label{fig:blast-radius}
\end{figure}

\subsection{Sensitivity studies}
\label{subsec:sens}
\daiyaan{@bhargava @ahmet please check if this sounds right. [BG]: Done. Took a pass} Power boosting allows partially failed NVL domains to operate at higher throughput (in order to not bottleneck healthy domains) in our design. One could potentially operate the healthy domains at a higher power to boost their throughput as well. Unfortunately, operating at higher powers comes with a tradeoff as it reduces performance/watt, as well as increase the datacenter power requirements. For the TP30 configuration in table \ref{tab:setup}, at 1.1$\times$ the baseline power performance/watt reduces by 2.8\% , and at 1.2$\times$ it reduces by 6.5\%. This may not be desirable in general.  However, since NTP-PW only boosts unhealthy racks, this reduction in performance/watt is only incurred by ~12\% of NVL domains in the worst-case. Furthermore, this doesn't increase the provisioning requirements as we repurpose the power from failed GPUs.  

In most of our experiments, we assume that a single GPU failure takes out only one GPU out of an  NVL domain. \cite{gpu-fail-char} found that 91\% of failures are uncontained memory errors or MMU errors, which only affect a single GPU; they also found that 5\% of errors are NVLink errors, which can propagate to other GPU. Additionally, in some architecutures it is simpler to discard a group of GPUs than a single GPU. In GB200-NVL72, there are 36 CPUs and 72 GPUs, and it is often simpler to discard an entire node with any GPU failures rather than try to operate with partially failed nodes. In Figure \ref{fig:blast-radius}, we vary the failure blast radius (number of GPUs taken out by a single GPU failure) and study its effect on NTP. Increasing blast-radius has no effect on DP-DROP because the effective blast-radius (due to dropping and entire DP replica) is already very high. For NTP and NTP-PW, we do see an increase in lost-throughput at larger blast radii, but both methods still outperform DP-DROP substantially. 

\begin{figure}[H]
    \centering
    \begin{subfigure}{0.47\columnwidth}
        \centering
        \includegraphics[width=\linewidth]{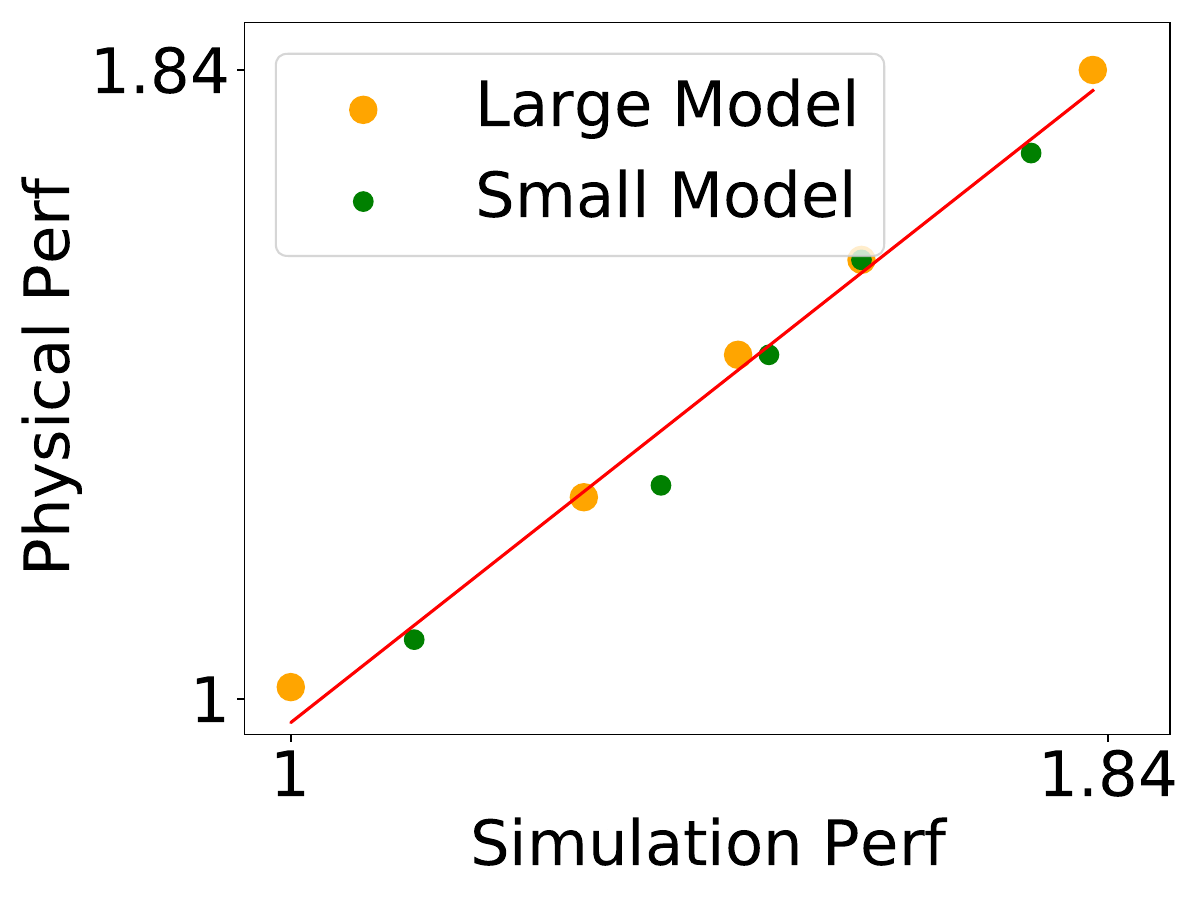}
        \caption{}
        \label{fig:sim-pow}
    \end{subfigure}
    \hfill
    \begin{subfigure}{0.47\columnwidth}
        \centering
        \includegraphics[width=\linewidth]{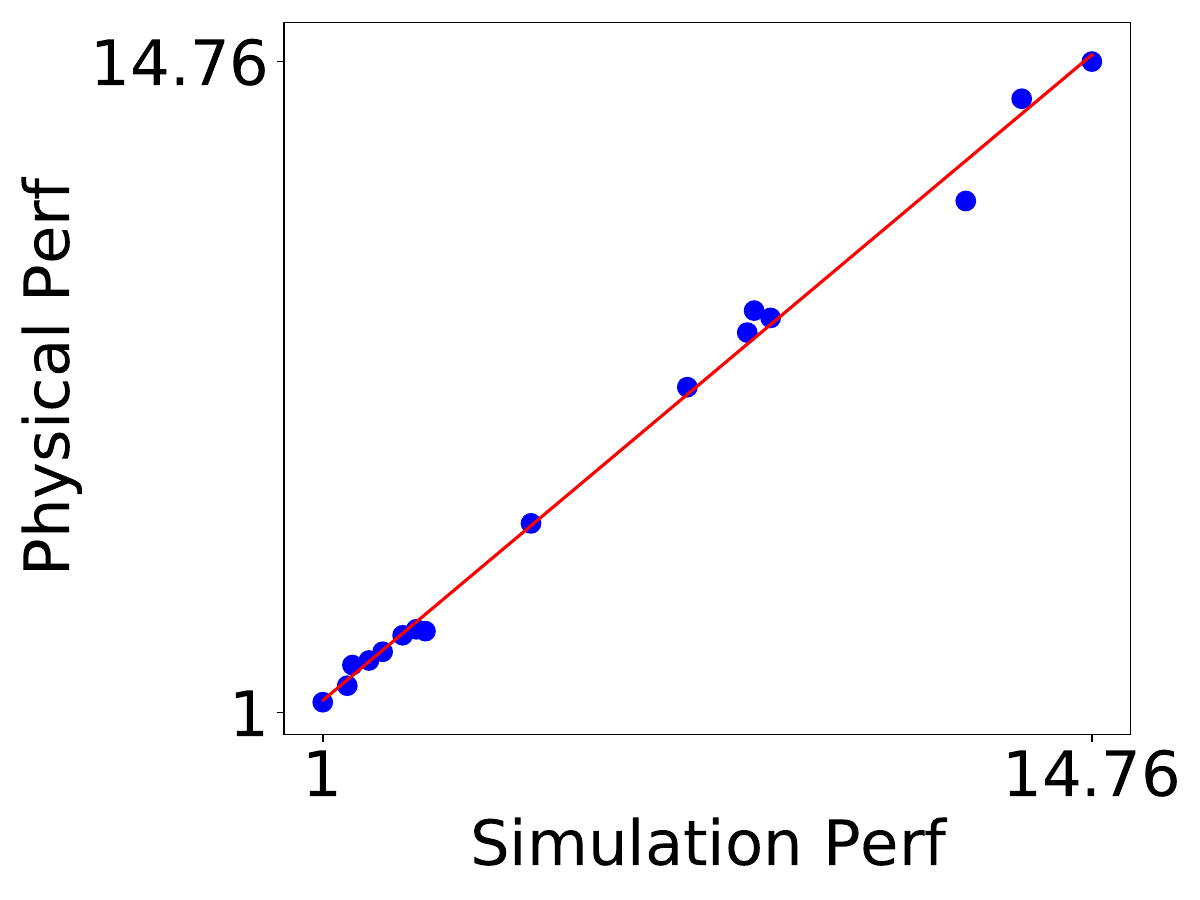}
        \caption{}
        \label{fig:sim-scale}
    \end{subfigure}
    \caption{Simulator shows high correlation with observed performance across different power levels and for multiple workloads at different scales (a), as well as for many different pretraining workloads (model, size, sequence length, scale) (b).}
    \label{fig:sim-combined}
\end{figure}

\daiyaan{regrouping + sharing results}

% \begin{figure}[H]
%     \centering
%     \includegraphics[width=\linewidth]{plots/sens/blast_radius.pdf}
%     \caption{We plot the fraction of total cluster GPUs lost due to failures at various failure blast-radii. Larger blast-radii reduce NTP's performance, but it still outperforms DP-DROP.}
%     \label{fig:blast-radius}
% \end{figure}
\section{Related Work}
\label{sec:related}

\noindent\textbf{Resilient DNN Training} Traditionally, DNN training has relied on checkpointing for resilience to failures \cite{mohan2021checkfreq, eisenman2022checknrun, wang2024fastpersist}; most recently in-memory checkpointing has been proposed \cite{wang2023gemini, wan2024bytecheckpoint} to reduce the overhead of both checkpointing as well as restarting from a checkpoint. To continue training from a checckpoint at a fixed batch size in the face of hardware failures requires spare GPUs to replace the failed GPUs. 

Recently, there has been interest in reconfiguring distributed DNN training jobs to continue training at a fixed batch size in settings where spare GPUs are unavailable or undesirable. Many works have been proposed that solve this problem for PP \cite{jang2023oobleck,recycle,thorpe2023bamboo}. Unfortunately, none of these works address the issue of \NVLdomain failure amplification, instead opting to drop an entire \NVLdomain / TP group.

\noindent\textbf{Overclocking} \cite{jalili2021cost} proposed immersion-cooling datacenters processors (including CPUs and GPUs) in order to be able to sustain higher clocks for cost and performance purposes. Among the proposed uses cases were VM buffer and crisis capacity reduction, highly related to the reduction of spare GPU capacity for maintaining training performance and batch size. \cite{patel2024characterizing} did an in-depth analysis of power-usage in LLM training and found that LLM training often reaches or exceeds GPU TDPs, showing that there is headroom to overclock GPUs with sufficient cooling capacity.

\noindent \textbf{DNN Training on Heterogeneous Resources} Resilient training is one form of heterogeneous training. With the increased cadence of introduction of new hardware, its likely that data-centers will house multiple different platform possibly across multiple generations. This requires large scale training jobs to deal with hardware heterogeneity, since these can be significantly different in terms of scale-up domain size, memory capacity, flops etc. Prior works have explored heterogeneous training with different GPU types\cite{metis_atc, sdpipe_vldb, hetpipe_atc, hap_eurosys}. NTP enables efficient training with heterogeneous scale-up domain sizes (as a result of failures reducing some scale-up domain sizes), particularly for jobs using large TP sizes. To our knowledge, we are the first to address this problem, and the first to propose hardware overclocking as part of our solution. 
\section{Conclusion}

Nonuniform Tensor Parallelism (NTP) can speed up LLM training by enabling use of larger scale-up domains while mitigating GPU time losses when some of the 1000s of GPUs inevitably fail.
With NTP, lost training throughput can be nearly as low as the percentage of failed GPUs, rather than 10 to 100$\times$ higher with traditional parallelization.
Furthermore, combined with new power boost support for a small fraction of non-failed GPUs, NTP could almost eliminate throughput loss.
%with $<$4\% increase in energy usage in the worst-case.
As a result, NTP will be important for more efficient LLM training on future hardware.

\bibliographystyle{ACM-Reference-Format}
\bibliography{ref}


\begin{thebibliography}{37}

%%% the Web address.

\ifx \showCODEN    \undefined \def \showCODEN     #1{\unskip}     \fi
\ifx \showISBNx    \undefined \def \showISBNx     #1{\unskip}     \fi
\ifx \showISBNxiii \undefined \def \showISBNxiii  #1{\unskip}     \fi
\ifx \showISSN     \undefined \def \showISSN      #1{\unskip}     \fi
\ifx \showLCCN     \undefined \def \showLCCN      #1{\unskip}     \fi
\ifx \shownote     \undefined \def \shownote      #1{#1}          \fi
\ifx \showarticletitle \undefined \def \showarticletitle #1{#1}   \fi
\ifx \showURL      \undefined \def \showURL       {\relax}        \fi
% invisible to TeX
\providecommand\bibfield[2]{#2}
\providecommand\bibinfo[2]{#2}
\providecommand\natexlab[1]{#1}
\providecommand\showeprint[2][]{arXiv:#2}

\bibitem[AMD(2025)]%
        {infinity-fabric}
\bibfield{author}{\bibinfo{person}{AMD}.} \bibinfo{year}{2025}\natexlab{}.
\newblock \bibinfo{title}{AMD Infinity Fabric Link}.
\newblock
\urldef\tempurl%
\url{https://www.amd.com/content/dam/amd/en/documents/instinct-tech-docs/other/56978.pdf}
\showURL{%
\tempurl}


\bibitem[Brown et~al\mbox{.}(2020)]%
        {gpt3}
\bibfield{author}{\bibinfo{person}{Tom~B. Brown}, \bibinfo{person}{Benjamin Mann}, \bibinfo{person}{Nick Ryder}, \bibinfo{person}{Melanie Subbiah}, \bibinfo{person}{Jared Kaplan}, \bibinfo{person}{Prafulla Dhariwal}, \bibinfo{person}{Arvind Neelakantan}, \bibinfo{person}{Pranav Shyam}, \bibinfo{person}{Girish Sastry}, \bibinfo{person}{Amanda Askell}, \bibinfo{person}{Sandhini Agarwal}, \bibinfo{person}{Ariel Herbert-Voss}, \bibinfo{person}{Gretchen Krueger}, \bibinfo{person}{Tom Henighan}, \bibinfo{person}{Rewon Child}, \bibinfo{person}{Aditya Ramesh}, \bibinfo{person}{Daniel~M. Ziegler}, \bibinfo{person}{Jeffrey Wu}, \bibinfo{person}{Clemens Winter}, \bibinfo{person}{Christopher Hesse}, \bibinfo{person}{Mark Chen}, \bibinfo{person}{Eric Sigler}, \bibinfo{person}{Mateusz Litwin}, \bibinfo{person}{Scott Gray}, \bibinfo{person}{Benjamin Chess}, \bibinfo{person}{Jack Clark}, \bibinfo{person}{Christopher Berner}, \bibinfo{person}{Sam McCandlish}, \bibinfo{person}{Alec Radford}, \bibinfo{person}{Ilya Sutskever},
  {and} \bibinfo{person}{Dario Amodei}.} \bibinfo{year}{2020}\natexlab{}.
\newblock \showarticletitle{Language Models are Few-Shot Learners}.
\newblock \bibinfo{journal}{\emph{arXiv preprint arXiv:2005.14165}} (\bibinfo{year}{2020}).
\newblock
\urldef\tempurl%
\url{https://arxiv.org/abs/2005.14165}
\showURL{%
\tempurl}


\bibitem[Cui et~al\mbox{.}(2025)]%
        {gpu-fail-char}
\bibfield{author}{\bibinfo{person}{Shengkun Cui}, \bibinfo{person}{Archit Patke}, \bibinfo{person}{Ziheng Chen}, \bibinfo{person}{Aditya Ranjan}, \bibinfo{person}{Hung Nguyen}, \bibinfo{person}{Phuong Cao}, \bibinfo{person}{Saurabh Jha}, \bibinfo{person}{Brett Bode}, \bibinfo{person}{Gregory Bauer}, \bibinfo{person}{Chandra Narayanaswami}, \bibinfo{person}{Daby Sow}, \bibinfo{person}{Catello~Di Martino}, \bibinfo{person}{Zbigniew~T. Kalbarczyk}, {and} \bibinfo{person}{Ravishankar~K. Iyer}.} \bibinfo{year}{2025}\natexlab{}.
\newblock \bibinfo{title}{Characterizing GPU Resilience and Impact on AI/HPC Systems}.
\newblock
\showeprint[arxiv]{2503.11901}~[cs.DC]
\urldef\tempurl%
\url{https://arxiv.org/abs/2503.11901}
\showURL{%
\tempurl}


\bibitem[Eisenman et~al\mbox{.}(2022)]%
        {eisenman2022checknrun}
\bibfield{author}{\bibinfo{person}{Assaf Eisenman}, \bibinfo{person}{Kiran~Kumar Matam}, \bibinfo{person}{Steven Ingram}, \bibinfo{person}{Dheevatsa Mudigere}, \bibinfo{person}{Raghuraman Krishnamoorthi}, \bibinfo{person}{Krishnakumar Nair}, \bibinfo{person}{Misha Smelyanskiy}, {and} \bibinfo{person}{Murali Annavaram}.} \bibinfo{year}{2022}\natexlab{}.
\newblock \showarticletitle{Check-N-Run: A Checkpointing System for Training Deep Learning Recommendation Models}. In \bibinfo{booktitle}{\emph{19th USENIX Symposium on Networked Systems Design and Implementation (NSDI)}}. \bibinfo{pages}{323--338}.
\newblock
\urldef\tempurl%
\url{https://www.usenix.org/system/files/nsdi22-paper-eisenman.pdf}
\showURL{%
\tempurl}


\bibitem[Fan et~al\mbox{.}(2024)]%
        {trainium}
\bibfield{author}{\bibinfo{person}{Haozheng Fan}, \bibinfo{person}{Hao Zhou}, \bibinfo{person}{Guangtai Huang}, \bibinfo{person}{Parameswaran Raman}, \bibinfo{person}{Xinwei Fu}, \bibinfo{person}{Gaurav Gupta}, \bibinfo{person}{Dhananjay Ram}, \bibinfo{person}{Yida Wang}, {and} \bibinfo{person}{Jun Huan}.} \bibinfo{year}{2024}\natexlab{}.
\newblock \showarticletitle{{HLAT}: High-quality Large Language Model Pre-trained on {AWS} Trainium}.
\newblock \bibinfo{journal}{\emph{arXiv preprint arXiv:2404.10630}} (\bibinfo{year}{2024}).
\newblock
\urldef\tempurl%
\url{https://arxiv.org/abs/2404.10630}
\showURL{%
\tempurl}


\bibitem[Fedus et~al\mbox{.}(2022)]%
        {switch_transformers}
\bibfield{author}{\bibinfo{person}{William Fedus}, \bibinfo{person}{Barret Zoph}, {and} \bibinfo{person}{Noam Shazeer}.} \bibinfo{year}{2022}\natexlab{}.
\newblock \showarticletitle{Switch Transformers: Scaling to Trillion Parameter Models with Simple and Efficient Sparsity}.
\newblock \bibinfo{journal}{\emph{Journal of Machine Learning Research}}  \bibinfo{volume}{23} (\bibinfo{year}{2022}), \bibinfo{pages}{1--40}.
\newblock
\urldef\tempurl%
\url{https://arxiv.org/abs/2101.03961}
\showURL{%
\tempurl}


\bibitem[Gandhi et~al\mbox{.}(2024)]%
        {recycle}
\bibfield{author}{\bibinfo{person}{Swapnil Gandhi}, \bibinfo{person}{Mark Zhao}, \bibinfo{person}{Athinagoras Skiadopoulos}, {and} \bibinfo{person}{Christos Kozyrakis}.} \bibinfo{year}{2024}\natexlab{}.
\newblock \showarticletitle{ReCycle: Resilient Training of Large DNNs using Pipeline Adaptation}. In \bibinfo{booktitle}{\emph{Proceedings of the ACM SIGOPS 30th Symposium on Operating Systems Principles}} (Austin, TX, USA) \emph{(\bibinfo{series}{SOSP '24})}. \bibinfo{publisher}{Association for Computing Machinery}, \bibinfo{address}{New York, NY, USA}, \bibinfo{pages}{211–228}.
\newblock
\showISBNx{9798400712517}
\href{https://doi.org/10.1145/3694715.3695960}{doi:\nolinkurl{10.1145/3694715.3695960}}


\bibitem[Goldwasser et~al\mbox{.}(2024)]%
        {gb200-nvl72}
\bibfield{author}{\bibinfo{person}{Ivan Goldwasser}, \bibinfo{person}{Harry Petty}, \bibinfo{person}{Pradyumna Desale}, {and} \bibinfo{person}{Kirthi Devleker}.} \bibinfo{year}{2024}\natexlab{}.
\newblock \bibinfo{title}{{NVIDIA GB200 NVL72 Delivers Trillion-Parameter LLM Training and Real-Time Inference}}.
\newblock
\urldef\tempurl%
\url{https://developer.nvidia.com/blog/nvidia-gb200-nvl72-delivers-trillion-parameter-llm-training-and-real-time-inference/}
\showURL{%
\tempurl}
\newblock
\shownote{Accessed: 2025-03-12}.


\bibitem[Google(2025)]%
        {tpu-pod}
\bibfield{author}{\bibinfo{person}{Google}.} \bibinfo{year}{2025}\natexlab{}.
\newblock \bibinfo{title}{TPU Architecture (TPU Pod section)}.
\newblock
\urldef\tempurl%
\url{https://cloud.google.com/tpu/docs/system-architecture-tpu-vm#tpu-pod}
\showURL{%
\tempurl}


\bibitem[Harlap et~al\mbox{.}(2019)]%
        {pipedream}
\bibfield{author}{\bibinfo{person}{Aaron Harlap}, \bibinfo{person}{Deepak Narayanan}, \bibinfo{person}{Amar Phanishayee}, \bibinfo{person}{Vivek Seshadri}, \bibinfo{person}{Nikhil Devanur}, \bibinfo{person}{Greg Ganger}, {and} \bibinfo{person}{Phil Gibbons}.} \bibinfo{year}{2019}\natexlab{}.
\newblock \showarticletitle{PipeDream: Fast and Efficient Pipeline Parallel DNN Training}. In \bibinfo{booktitle}{\emph{Proceedings of the 27th ACM Symposium on Operating Systems Principles (SOSP)}}.
\newblock
\urldef\tempurl%
\url{https://arxiv.org/abs/1806.03377}
\showURL{%
\tempurl}


\bibitem[Huang et~al\mbox{.}(2019)]%
        {gpipe}
\bibfield{author}{\bibinfo{person}{Yanping Huang}, \bibinfo{person}{Youlong Cheng}, \bibinfo{person}{Ankur Bapna}, \bibinfo{person}{Orhan Firat}, \bibinfo{person}{Mia~Xu Chen}, \bibinfo{person}{Dehao Chen}, \bibinfo{person}{HyoukJoong Lee}, \bibinfo{person}{Jiquan Ngiam}, \bibinfo{person}{Quoc~V. Le}, \bibinfo{person}{Yonghui Wu}, {and} \bibinfo{person}{Zhifeng Chen}.} \bibinfo{year}{2019}\natexlab{}.
\newblock \showarticletitle{GPipe: Easy Scaling with Micro-Batch Pipeline Parallelism}. In \bibinfo{booktitle}{\emph{Advances in Neural Information Processing Systems (NeurIPS)}}.
\newblock
\urldef\tempurl%
\url{https://arxiv.org/abs/1811.06965}
\showURL{%
\tempurl}


\bibitem[Jalili et~al\mbox{.}(2021)]%
        {jalili2021cost}
\bibfield{author}{\bibinfo{person}{Majid Jalili}, \bibinfo{person}{Ioannis Manousakis}, \bibinfo{person}{{\'I}{\~n}igo Goiri}, \bibinfo{person}{Pulkit~A Misra}, \bibinfo{person}{Ashish Raniwala}, \bibinfo{person}{Husam Alissa}, \bibinfo{person}{Bharath Ramakrishnan}, \bibinfo{person}{Phillip Tuma}, \bibinfo{person}{Christian Belady}, \bibinfo{person}{Marcus Fontoura}, {et~al\mbox{.}}} \bibinfo{year}{2021}\natexlab{}.
\newblock \showarticletitle{Cost-efficient overclocking in immersion-cooled datacenters}. In \bibinfo{booktitle}{\emph{2021 ACM/IEEE 48th Annual International Symposium on Computer Architecture (ISCA)}}. IEEE, \bibinfo{pages}{623--636}.
\newblock


\bibitem[Jang et~al\mbox{.}(2023)]%
        {jang2023oobleck}
\bibfield{author}{\bibinfo{person}{Insu Jang}, \bibinfo{person}{Zhenning Yang}, \bibinfo{person}{Zhen Zhang}, \bibinfo{person}{Xin Jin}, {and} \bibinfo{person}{Mosharaf Chowdhury}.} \bibinfo{year}{2023}\natexlab{}.
\newblock \showarticletitle{Oobleck: Resilient Distributed Training of Large Models Using Pipeline Templates}. In \bibinfo{booktitle}{\emph{Proceedings of the ACM SIGOPS 29th Symposium on Operating Systems Principles (SOSP)}}. \bibinfo{pages}{362--377}.
\newblock
\href{https://doi.org/10.1145/3600006.3613152}{doi:\nolinkurl{10.1145/3600006.3613152}}


\bibitem[Jouppi et~al\mbox{.}(2023)]%
        {tpuv4}
\bibfield{author}{\bibinfo{person}{Norman~P. Jouppi}, \bibinfo{person}{George Kurian}, \bibinfo{person}{Sheng Li}, \bibinfo{person}{Peter Ma}, \bibinfo{person}{Rahul Nagarajan}, \bibinfo{person}{Lifeng Nai}, \bibinfo{person}{Nishant Patil}, \bibinfo{person}{Suvinay Subramanian}, \bibinfo{person}{Andy Swing}, \bibinfo{person}{Brian Towles}, \bibinfo{person}{Cliff Young}, \bibinfo{person}{Xiang Zhou}, \bibinfo{person}{Zongwei Zhou}, {and} \bibinfo{person}{David Patterson}.} \bibinfo{year}{2023}\natexlab{}.
\newblock \showarticletitle{TPU v4: An Optically Reconfigurable Supercomputer for Machine Learning with Hardware Support for Embeddings}. In \bibinfo{booktitle}{\emph{Proceedings of the 50th Annual International Symposium on Computer Architecture (ISCA)}}.
\newblock
\urldef\tempurl%
\url{https://arxiv.org/abs/2304.01433}
\showURL{%
\tempurl}


\bibitem[Kokolis et~al\mbox{.}(2025)]%
        {metafail}
\bibfield{author}{\bibinfo{person}{Apostolos Kokolis}, \bibinfo{person}{Michael Kuchnik}, \bibinfo{person}{John Hoffman}, \bibinfo{person}{Adithya Kumar}, \bibinfo{person}{Parth Malani}, \bibinfo{person}{Faye Ma}, \bibinfo{person}{Zachary DeVito}, \bibinfo{person}{Shubho Sengupta}, \bibinfo{person}{Kalyan Saladi}, {and} \bibinfo{person}{Carole-Jean Wu}.} \bibinfo{year}{2025}\natexlab{}.
\newblock \bibinfo{title}{Revisiting Reliability in Large-Scale Machine Learning Research Clusters}.
\newblock
\showeprint[arxiv]{2410.21680}~[cs.DC]
\urldef\tempurl%
\url{https://arxiv.org/abs/2410.21680}
\showURL{%
\tempurl}


\bibitem[Li et~al\mbox{.}(2020)]%
        {pytorch_distributed}
\bibfield{author}{\bibinfo{person}{Shen Li}, \bibinfo{person}{Yanli Zhao}, \bibinfo{person}{Rohan Varma}, \bibinfo{person}{Omkar Salpekar}, \bibinfo{person}{Pieter Noordhuis}, \bibinfo{person}{Teng Li}, \bibinfo{person}{Adam Paszke}, \bibinfo{person}{Jeff Smith}, \bibinfo{person}{Brian Vaughan}, \bibinfo{person}{Pritam Damania}, {and} \bibinfo{person}{Soumith Chintala}.} \bibinfo{year}{2020}\natexlab{}.
\newblock \showarticletitle{PyTorch Distributed: Experiences on Accelerating Data Parallel Training}.
\newblock \bibinfo{journal}{\emph{arXiv preprint arXiv:2006.15704}} (\bibinfo{year}{2020}).
\newblock
\urldef\tempurl%
\url{https://arxiv.org/abs/2006.15704}
\showURL{%
\tempurl}


\bibitem[Liu et~al\mbox{.}(2023)]%
        {ring_attention}
\bibfield{author}{\bibinfo{person}{Hao Liu}, \bibinfo{person}{Matei Zaharia}, {and} \bibinfo{person}{Pieter Abbeel}.} \bibinfo{year}{2023}\natexlab{}.
\newblock \showarticletitle{Ring Attention with Blockwise Transformers for Near-Infinite Context}. In \bibinfo{booktitle}{\emph{Advances in Neural Information Processing Systems (NeurIPS)}}.
\newblock
\urldef\tempurl%
\url{https://arxiv.org/abs/2310.01889}
\showURL{%
\tempurl}


\bibitem[{Llama Team, AI @ Meta}(2024)]%
        {llama3}
\bibfield{author}{\bibinfo{person}{{Llama Team, AI @ Meta}}.} \bibinfo{year}{2024}\natexlab{}.
\newblock \showarticletitle{The Llama 3 Herd of Models}.
\newblock \bibinfo{journal}{\emph{arXiv preprint arXiv:2407.21783}} (\bibinfo{year}{2024}).
\newblock
\urldef\tempurl%
\url{https://arxiv.org/abs/2407.21783}
\showURL{%
\tempurl}
\newblock
\shownote{A detailed contributor list can be found in the appendix of this paper.}.


\bibitem[Miao et~al\mbox{.}(2023)]%
        {sdpipe_vldb}
\bibfield{author}{\bibinfo{person}{Xupeng Miao}, \bibinfo{person}{Yining Shi}, \bibinfo{person}{Zhi Yang}, \bibinfo{person}{Bin Cui}, {and} \bibinfo{person}{Zhihao Jia}.} \bibinfo{year}{2023}\natexlab{}.
\newblock \showarticletitle{SDPipe: A Semi-Decentralized Framework for Heterogeneity-Aware Pipeline-parallel Training}.
\newblock \bibinfo{journal}{\emph{Proc. VLDB Endow.}} \bibinfo{volume}{16}, \bibinfo{number}{9} (\bibinfo{date}{May} \bibinfo{year}{2023}), \bibinfo{pages}{2354–2363}.
\newblock
\showISSN{2150-8097}
\href{https://doi.org/10.14778/3598581.3598604}{doi:\nolinkurl{10.14778/3598581.3598604}}


\bibitem[Mohan et~al\mbox{.}(2021)]%
        {mohan2021checkfreq}
\bibfield{author}{\bibinfo{person}{Jayashree Mohan}, \bibinfo{person}{Amar Phanishayee}, {and} \bibinfo{person}{Vijay Chidambaram}.} \bibinfo{year}{2021}\natexlab{}.
\newblock \showarticletitle{CheckFreq: Frequent, Fine-Grained DNN Checkpointing}. In \bibinfo{booktitle}{\emph{19th USENIX Conference on File and Storage Technologies (FAST)}}. \bibinfo{pages}{203--216}.
\newblock
\urldef\tempurl%
\url{https://www.usenix.org/conference/fast21/presentation/mohan}
\showURL{%
\tempurl}


\bibitem[Narayanan et~al\mbox{.}(2021)]%
        {megatron2}
\bibfield{author}{\bibinfo{person}{Deepak Narayanan}, \bibinfo{person}{Mohammad Shoeybi}, \bibinfo{person}{Jared Casper}, \bibinfo{person}{Patrick LeGresley}, \bibinfo{person}{Mostofa Patwary}, \bibinfo{person}{Vijay~Anand Korthikanti}, \bibinfo{person}{Dmitri Vainbrand}, \bibinfo{person}{Prethvi Kashinkunti}, \bibinfo{person}{Julie Bernauer}, \bibinfo{person}{Bryan Catanzaro}, \bibinfo{person}{Amar Phanishayee}, {and} \bibinfo{person}{Matei Zaharia}.} \bibinfo{year}{2021}\natexlab{}.
\newblock \showarticletitle{Efficient Large-Scale Language Model Training on GPU Clusters Using Megatron-LM}. In \bibinfo{booktitle}{\emph{Proceedings of the International Conference for High Performance Computing, Networking, Storage and Analysis (SC)}}.
\newblock
\urldef\tempurl%
\url{https://arxiv.org/abs/2104.04473}
\showURL{%
\tempurl}


\bibitem[NVIDIA(2024a)]%
        {mlperf-inference-nvidia}
\bibfield{author}{\bibinfo{person}{NVIDIA}.} \bibinfo{year}{2024}\natexlab{a}.
\newblock \bibinfo{title}{Accelerating {LLM} Inference with {TensorRT}-{LLM} and {MLPerf}}.
\newblock
\urldef\tempurl%
\url{https://blogs.nvidia.com/blog/tensorrt-llm-inference-mlperf/}
\showURL{%
\tempurl}
\newblock
\shownote{Accessed: 2025-03-12}.


\bibitem[NVIDIA(2024b)]%
        {megatron-framework}
\bibfield{author}{\bibinfo{person}{NVIDIA}.} \bibinfo{year}{2024}\natexlab{b}.
\newblock \bibinfo{title}{{Megatron-LM: Training Multi-Billion Parameter Language Models}}.
\newblock
\urldef\tempurl%
\url{https://github.com/NVIDIA/Megatron-LM}
\showURL{%
\tempurl}
\newblock
\shownote{Accessed: 2025-03-12}.


\bibitem[NVIDIA(2025)]%
        {nvlink}
\bibfield{author}{\bibinfo{person}{NVIDIA}.} \bibinfo{year}{2025}\natexlab{}.
\newblock \bibinfo{title}{NVLink and NVLink Switch}.
\newblock
\urldef\tempurl%
\url{https://www.nvidia.com/en-in/data-center/nvlink/}
\showURL{%
\tempurl}


\bibitem[Park et~al\mbox{.}(2020)]%
        {hetpipe_atc}
\bibfield{author}{\bibinfo{person}{Jay~H. Park}, \bibinfo{person}{Gyeongchan Yun}, \bibinfo{person}{Chang~M. Yi}, \bibinfo{person}{Nguyen~T. Nguyen}, \bibinfo{person}{Seungmin Lee}, \bibinfo{person}{Jaesik Choi}, \bibinfo{person}{Sam~H. Noh}, {and} \bibinfo{person}{Young ri Choi}.} \bibinfo{year}{2020}\natexlab{}.
\newblock \showarticletitle{{HetPipe}: Enabling Large {DNN} Training on (Whimpy) Heterogeneous {GPU} Clusters through Integration of Pipelined Model Parallelism and Data Parallelism}. In \bibinfo{booktitle}{\emph{2020 USENIX Annual Technical Conference (USENIX ATC 20)}}. \bibinfo{publisher}{USENIX Association}, \bibinfo{pages}{307--321}.
\newblock
\showISBNx{978-1-939133-14-4}
\urldef\tempurl%
\url{https://www.usenix.org/conference/atc20/presentation/park}
\showURL{%
\tempurl}


\bibitem[Patel et~al\mbox{.}(2024)]%
        {patel2024characterizing}
\bibfield{author}{\bibinfo{person}{Pratyush Patel}, \bibinfo{person}{Esha Choukse}, \bibinfo{person}{Chaojie Zhang}, \bibinfo{person}{{\'I}{\~n}igo Goiri}, \bibinfo{person}{Brijesh Warrier}, \bibinfo{person}{Nithish Mahalingam}, {and} \bibinfo{person}{Ricardo Bianchini}.} \bibinfo{year}{2024}\natexlab{}.
\newblock \showarticletitle{Characterizing power management opportunities for llms in the cloud}. In \bibinfo{booktitle}{\emph{Proceedings of the 29th ACM International Conference on Architectural Support for Programming Languages and Operating Systems, Volume 3}}. \bibinfo{pages}{207--222}.
\newblock


\bibitem[Rajbhandari et~al\mbox{.}(2020)]%
        {zero}
\bibfield{author}{\bibinfo{person}{Samyam Rajbhandari}, \bibinfo{person}{Jeff Rasley}, \bibinfo{person}{Olatunji Ruwase}, {and} \bibinfo{person}{Yuxiong He}.} \bibinfo{year}{2020}\natexlab{}.
\newblock \showarticletitle{ZeRO: Memory Optimizations Toward Training Trillion Parameter Models}. In \bibinfo{booktitle}{\emph{Proceedings of the International Conference for High Performance Computing, Networking, Storage and Analysis (SC)}}.
\newblock
\urldef\tempurl%
\url{https://arxiv.org/abs/1910.02054}
\showURL{%
\tempurl}


\bibitem[Shoeybi et~al\mbox{.}(2019)]%
        {megatron1}
\bibfield{author}{\bibinfo{person}{Mohammad Shoeybi}, \bibinfo{person}{Mostofa Patwary}, \bibinfo{person}{Raul Puri}, \bibinfo{person}{Patrick LeGresley}, \bibinfo{person}{Jared Casper}, {and} \bibinfo{person}{Bryan Catanzaro}.} \bibinfo{year}{2019}\natexlab{}.
\newblock \showarticletitle{Megatron-LM: Training Multi-Billion Parameter Language Models Using Model Parallelism}.
\newblock \bibinfo{journal}{\emph{arXiv preprint arXiv:1909.08053}} (\bibinfo{year}{2019}).
\newblock
\urldef\tempurl%
\url{https://arxiv.org/abs/1909.08053}
\showURL{%
\tempurl}


\bibitem[Systems(2021)]%
        {cerebras}
\bibfield{author}{\bibinfo{person}{Cerebras Systems}.} \bibinfo{year}{2021}\natexlab{}.
\newblock \bibinfo{title}{CS-2 Data Sheet}.
\newblock
\urldef\tempurl%
\url{https://f.hubspotusercontent30.net/hubfs/8968533/CS-2%20Data%20Sheet.pdf}
\showURL{%
\tempurl}
\newblock
\shownote{Accessed: 2025-03-12}.


\bibitem[Team(2024)]%
        {gemini}
\bibfield{author}{\bibinfo{person}{Gemini Team}.} \bibinfo{year}{2024}\natexlab{}.
\newblock \bibinfo{title}{Gemini: A Family of Highly Capable Multimodal Models}.
\newblock
\showeprint[arxiv]{2312.11805}~[cs.CL]
\urldef\tempurl%
\url{https://arxiv.org/abs/2312.11805}
\showURL{%
\tempurl}


\bibitem[Thorpe et~al\mbox{.}(2023)]%
        {thorpe2023bamboo}
\bibfield{author}{\bibinfo{person}{John Thorpe}, \bibinfo{person}{Pengzhan Zhao}, \bibinfo{person}{Jonathan Eyolfson}, \bibinfo{person}{Yifan Qiao}, \bibinfo{person}{Zhihao Jia}, \bibinfo{person}{Minjia Zhang}, \bibinfo{person}{Ravi Netravali}, {and} \bibinfo{person}{Guoqing~Harry Xu}.} \bibinfo{year}{2023}\natexlab{}.
\newblock \showarticletitle{Bamboo: Making Preemptible Instances Resilient for Affordable Training of Large {DNNs}}. In \bibinfo{booktitle}{\emph{20th USENIX Symposium on Networked Systems Design and Implementation (NSDI 23)}}. \bibinfo{publisher}{USENIX Association}, \bibinfo{address}{Boston, MA}, \bibinfo{pages}{497--513}.
\newblock
\showISBNx{978-1-939133-33-5}
\urldef\tempurl%
\url{https://www.usenix.org/conference/nsdi23/presentation/thorpe}
\showURL{%
\tempurl}


\bibitem[Um et~al\mbox{.}(2024)]%
        {metis_atc}
\bibfield{author}{\bibinfo{person}{Taegeon Um}, \bibinfo{person}{Byungsoo Oh}, \bibinfo{person}{Minyoung Kang}, \bibinfo{person}{Woo-Yeon Lee}, \bibinfo{person}{Goeun Kim}, \bibinfo{person}{Dongseob Kim}, \bibinfo{person}{Youngtaek Kim}, \bibinfo{person}{Mohd Muzzammil}, {and} \bibinfo{person}{Myeongjae Jeon}.} \bibinfo{year}{2024}\natexlab{}.
\newblock \showarticletitle{Metis: Fast Automatic Distributed Training on Heterogeneous {GPUs}}. In \bibinfo{booktitle}{\emph{2024 USENIX Annual Technical Conference (USENIX ATC 24)}}. \bibinfo{publisher}{USENIX Association}, \bibinfo{address}{Santa Clara, CA}, \bibinfo{pages}{563--578}.
\newblock
\showISBNx{978-1-939133-41-0}
\urldef\tempurl%
\url{https://www.usenix.org/conference/atc24/presentation/um}
\showURL{%
\tempurl}


\bibitem[Wan et~al\mbox{.}(2024)]%
        {wan2024bytecheckpoint}
\bibfield{author}{\bibinfo{person}{Borui Wan}, \bibinfo{person}{Mingji Han}, \bibinfo{person}{Yiyao Sheng}, \bibinfo{person}{Zhichao Lai}, \bibinfo{person}{Mofan Zhang}, \bibinfo{person}{Junda Zhang}, \bibinfo{person}{Yanghua Peng}, \bibinfo{person}{Haibin Lin}, \bibinfo{person}{Xin Liu}, {and} \bibinfo{person}{Chuan Wu}.} \bibinfo{year}{2024}\natexlab{}.
\newblock \showarticletitle{ByteCheckpoint: A Unified Checkpointing System for LLM Development}.
\newblock \bibinfo{journal}{\emph{arXiv preprint arXiv:2407.20143}} (\bibinfo{year}{2024}).
\newblock
\urldef\tempurl%
\url{https://arxiv.org/abs/2407.20143v1}
\showURL{%
\tempurl}


\bibitem[Wang et~al\mbox{.}(2024)]%
        {wang2024fastpersist}
\bibfield{author}{\bibinfo{person}{Guanhua Wang}, \bibinfo{person}{Olatunji Ruwase}, \bibinfo{person}{Bing Xie}, {and} \bibinfo{person}{Yuxiong He}.} \bibinfo{year}{2024}\natexlab{}.
\newblock \showarticletitle{FastPersist: Accelerating Model Checkpointing in Deep Learning}.
\newblock \bibinfo{journal}{\emph{arXiv preprint arXiv:2406.13768}} (\bibinfo{year}{2024}).
\newblock
\urldef\tempurl%
\url{https://arxiv.org/abs/2406.13768}
\showURL{%
\tempurl}


\bibitem[Wang et~al\mbox{.}(2023)]%
        {wang2023gemini}
\bibfield{author}{\bibinfo{person}{Zhuang Wang}, \bibinfo{person}{Zhen Jia}, \bibinfo{person}{Shuai Zheng}, \bibinfo{person}{Zhen Zhang}, \bibinfo{person}{Xinwei Fu}, \bibinfo{person}{T.~S.~Eugene Ng}, {and} \bibinfo{person}{Yida Wang}.} \bibinfo{year}{2023}\natexlab{}.
\newblock \showarticletitle{GEMINI: Fast Failure Recovery in Distributed Training with In-Memory Checkpoints}. In \bibinfo{booktitle}{\emph{Proceedings of the 29th Symposium on Operating Systems Principles}} (Koblenz, Germany) \emph{(\bibinfo{series}{SOSP '23})}. \bibinfo{publisher}{Association for Computing Machinery}, \bibinfo{address}{New York, NY, USA}, \bibinfo{pages}{364–381}.
\newblock
\showISBNx{9798400702297}
\href{https://doi.org/10.1145/3600006.3613145}{doi:\nolinkurl{10.1145/3600006.3613145}}


\bibitem[Wu et~al\mbox{.}(2016)]%
        {dynamo}
\bibfield{author}{\bibinfo{person}{Qiang Wu}, \bibinfo{person}{Qingyuan Deng}, \bibinfo{person}{Lakshmi Ganesh}, \bibinfo{person}{Chang-Hong Hsu}, \bibinfo{person}{Yun Jin}, \bibinfo{person}{Sanjeev Kumar}, \bibinfo{person}{Bin Li}, \bibinfo{person}{Justin Meza}, {and} \bibinfo{person}{Yee~Jiun Song}.} \bibinfo{year}{2016}\natexlab{}.
\newblock \showarticletitle{Dynamo: Facebook's Data Center-Wide Power Management System}. In \bibinfo{booktitle}{\emph{2016 ACM/IEEE 43rd Annual International Symposium on Computer Architecture (ISCA)}}. \bibinfo{pages}{469--480}.
\newblock
\href{https://doi.org/10.1109/ISCA.2016.48}{doi:\nolinkurl{10.1109/ISCA.2016.48}}


\bibitem[Zhang et~al\mbox{.}(2024)]%
        {hap_eurosys}
\bibfield{author}{\bibinfo{person}{Shiwei Zhang}, \bibinfo{person}{Lansong Diao}, \bibinfo{person}{Chuan Wu}, \bibinfo{person}{Zongyan Cao}, \bibinfo{person}{Siyu Wang}, {and} \bibinfo{person}{Wei Lin}.} \bibinfo{year}{2024}\natexlab{}.
\newblock \showarticletitle{HAP: SPMD DNN Training on Heterogeneous GPU Clusters with Automated Program Synthesis}. In \bibinfo{booktitle}{\emph{Proceedings of the Nineteenth European Conference on Computer Systems}} (Athens, Greece) \emph{(\bibinfo{series}{EuroSys '24})}. \bibinfo{publisher}{Association for Computing Machinery}, \bibinfo{address}{New York, NY, USA}, \bibinfo{pages}{524–541}.
\newblock
\showISBNx{9798400704376}
\href{https://doi.org/10.1145/3627703.3629580}{doi:\nolinkurl{10.1145/3627703.3629580}}


\end{thebibliography}

\clearpage

\appendix
\newpage
\section{Appendix}
\noindent\textbf{Implementation Details}
In Figure \ref{hook} we show a code snippet of the gradient-resharding hook in \ref{ntp-impl}. The resharding map (\textbf{recv\_splits} and \textbf{send\_splits}) are precomputed using Algorithm \ref{algo}. This hook is executed when the last gradient is computed and before the gradients are all-reduced.

\begin{figure}[h]
    \centering
\begin{lstlisting}[
        style=mystyle
    ]
def hook_full_tp(*unused):
    empty_shape = list(param.shape)
    empty_shape[param.partition_dim] = 0
    if tp_rank() < tp_base - tp_spares:
        # construct buffers for receiving grads
        # recv_splits precomputed using outputs of Algorithm 1
        input = [torch.empty(empty_shape) 
                 for _ in range(tp_world_size())]
        output = [torch.empty(empty_shape) 
                  for _ in range(tp_base - tp_spares)] 
        output += list(torch.split(
            param.extra_grad, param.recv_splits[tp_rank()]))
    else:
        # partition grads for sending to sync ranks
        # send_splits precomputed using outputs of Algorithm 1
        input = list(torch.split(
            param.grad, param.send_splits[tp_rank()]))
        output = [torch.empty(empty_shape) 
                  for _ in range(tp_world_size())]
    # reshard parameters using all-to-all
    torch.distributed.all_to_all(output, input)
\end{lstlisting}
    \caption{Code snippet for backward hook doing reshard before synchronization}
    \label{hook}
\end{figure}

In Figure \ref{post-sync-reshard} we show a code snippet of how the gradients are resharded after all-reduce. All buckets except for the last bucket will synchronize their gradients and wait for the last bucket. The last bucket will overlap the resharding of all previous buckets with its allreduce, and then reshard its own gradients.

\begin{figure}[h]
    \centering
\begin{lstlisting}[
        style=mystyle
    ]
def sync_reshard(bucket):
    if bucket.is_last:
        with torch.cuda.stream(reshard_stream):
            for syncd_bucket in self.buckets[:-1]:
                for param in syncd_bucket:
                    post_sync_reshard(param)

    for param in bucket:
        torch.distributed.allreduce(param)
        if bucket.is_last:
            post_sync_reshard(param)
\end{lstlisting}
    \caption{Code snippet showing gradient allreduce and post-sync reshard}
    \label{post-sync-reshard}
\end{figure}

\begin{figure}[H]
    \centering
    \includegraphics[width=\linewidth]{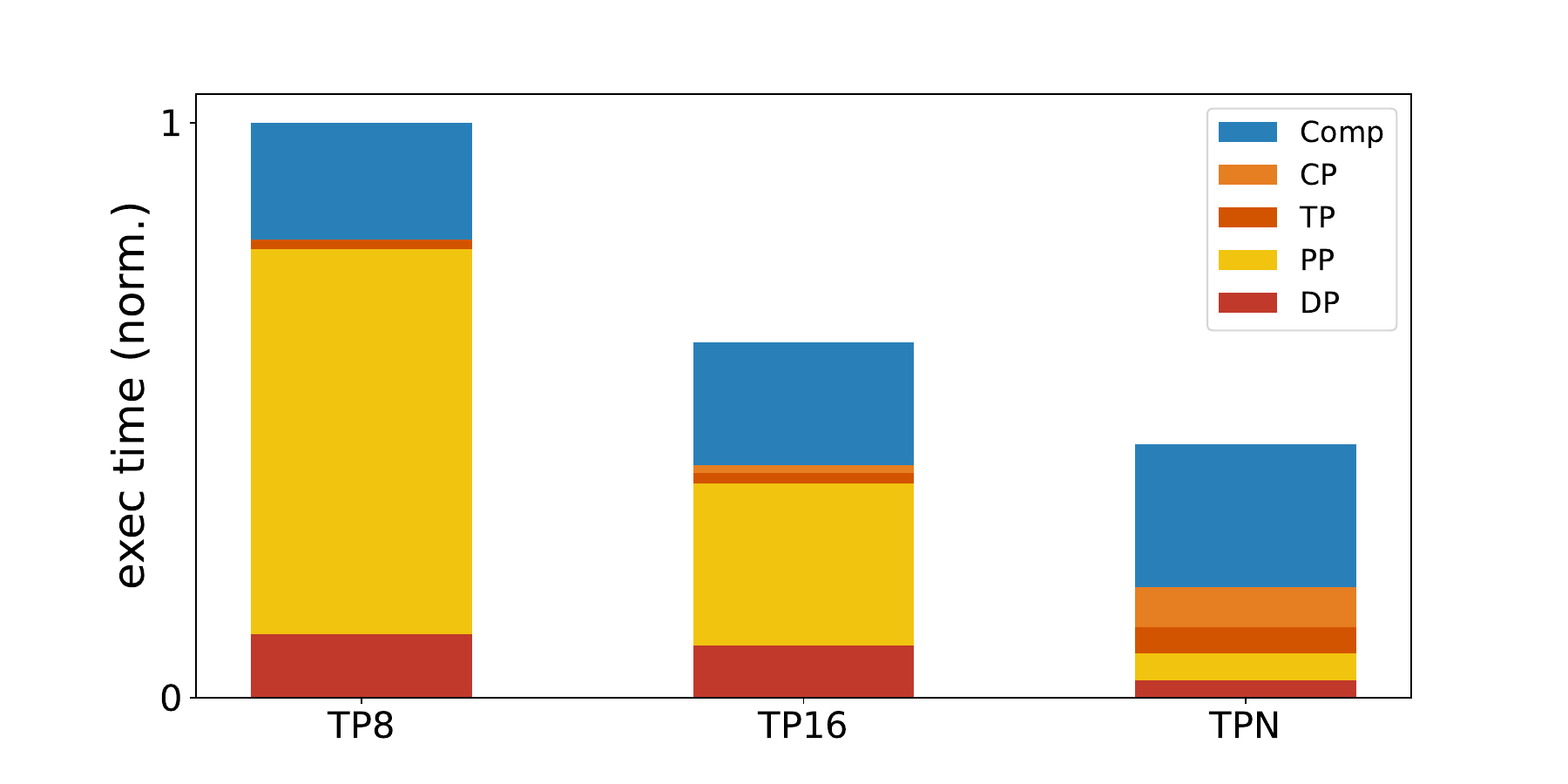}
    \caption{Execution time breakdown of different TP degree limits for 32K cluster with NVL32}
    \label{fig:dlsim_bar}
\end{figure}

\noindent \textbf{Communication overheads breakdown} In Figure \ref{fig:dlsim_bar} we show the execution time breakdown for 32K GPUs using NVL32 at different TP (Figure \ref{fig:motiv}b). When TP (and CP, which is folded into TP) are restricted, PP and DP are forced to expand. Due to the pipeline bubble issue, we see a huge portion of time spent on PP overhead. As TP/CP are allowed to increase, their overhead naturally increases, but the reduction in PP overhead more than makes up for it. In fact, even computation time goes up, but overall execution time sees a dramatic reduction. 

\end{document}